\documentclass[aps,11pt,axodraw,nofootinbib,superscriptaddress,aps]{revtex4}
\pdfoutput=1
\usepackage{epsfig}
\usepackage{amsmath}
\usepackage{bm}
\usepackage{times}
\usepackage{graphicx}
\usepackage{epstopdf}
\usepackage{amsfonts}
\usepackage{bm}
\usepackage{epsfig}
\usepackage{graphics}
\usepackage{xspace}
\usepackage[usenames]{color}

\def\C{\mathcal{C}}

\def\nn{\nonumber}
\def\bea{\begin{eqnarray}}
\def\eea{\end{eqnarray}}
\def\ba{\begin{eqnarray}}
\def\ea{\end{eqnarray}}
\def\be{\begin{equation}}
\def\ee{\end{equation}}

\def\beq{\begin{equation}}
\def\eeq{\end{equation}}

\def\nn{\nonumber}

\begin{document}

\title{\Large Fingerprinting Higgs Suspects at the LHC}
\author{J.R. Espinosa}
\affiliation{ICREA at IFAE, Universitat Aut{\`o}noma de Barcelona, 08193 Bellaterra, Barcelona, Spain}

\author{C. Grojean}
\affiliation{Theory Division, Physics Department, CERN, CH-1211 Geneva 23, Switzerland}

\author{M. M\"uhlleitner}
\affiliation{Institute for Theoretical Physics, Karlsruhe Institute of Technology, D-76128 Karlsruhe, Germany}

\author{M. Trott}
\affiliation{Theory Division, Physics Department, CERN, CH-1211 Geneva 23, Switzerland}

\date{\today}
\begin{abstract}
We outline a method for characterizing deviations from the properties
of a Standard Model (SM) Higgs boson. We apply it to current data in
order to characterize up to which degree the SM Higgs boson
interpretation is consistent with experiment.  We find that the 
SM Higgs boson is consistent with the current data set at the $82 \,
\%$ confidence level, based on data of excess events reported by CMS
and ATLAS, which are interpreted to be related to the mass scale $m_h
\sim 124-126 \, {\rm GeV}$, and on published $\rm CL_s$ exclusion regions.
We perform a global fit in terms of two parameters characterizing the deviation
from the SM value in the gauge and fermion couplings of a Higgs boson. We find two minima
in the global fit and identify observables that can remove this
degeneracy. An update for Moriond 2012 data is included in the Appendix, which finds that the SM Higgs boson is now
consistent with the current data set at only the $94 \,\%$ confidence level (which corresponds to $\sim 2 \, \sigma$ tension compared to the best fit point).
\end{abstract}
\maketitle
\section{Introduction}
The main goal of the LHC physics program is to unravel the origin of
electroweak symmetry breaking, with the leading explanation being the
Standard Model Higgs boson. In considering the SM Higgs hypothesis, or
more generally the correct description of electroweak symmetry
breaking (EWSB) as realized in nature, it is important to synthesize
the experimental results which are determined to date, in terms of
symmetries that must be (at least approximately) 
respected in a successful effective description of physics at the weak scale. 

Probes of the mechanism of electroweak symmetry breaking can be
summarized in terms of tests of the properties of the $\rm W^\pm$,
$\rm Z$ bosons through electroweak precision data observables (EWPD),
probes of the low energy flavour changing signatures of weak scale
physics, and direct searches. The results of the first two of these
probes indicate that a viable theory of EWSB can reduce parameter
tuning by adopting an approximate $\rm SU(2)_c$ custodial symmetry  to
comply with EWPD \cite{Susskind:1978ms,Weinberg:1979bn,Sikivie:1980hm}, and 
imposing approximate
Minimal Flavour Violation (MFV)
\cite{Chivukula:1987py,Hall:1990ac,D'Ambrosio:2002ex,Buras:2003jf,Cirigliano:2005ck}, 
in order to be robustly consistent with the results of the flavour physics program. The SM Higgs sector with renormalizable couplings to the fermions
is exactly consistent with MFV by definition, but it is only
approximately $\rm SU(2)_c$ symmetric. 

Recently, ATLAS and CMS have reported significant exclusion limits\footnote{These values for exclusions and the data used in the body of the paper correspond to pre-Moriond 2012 values.
The update of the fit for Moriond 2012 data is in Appendix B.} for
a Standard Model Higgs boson over a broad range of masses. 
The current exclusion regions reported by ATLAS
\cite{Collaboration:2012si} are $112.7 \leq m_h \leq115.5 \, {\rm
  GeV}$, $ 131 \leq m_h \leq 237  \, {\rm GeV}$ and  $251 \leq m_h
\leq 468  \, {\rm GeV}$ at $95 \%$ confidence level (CL). While the Higgs mass region excluded
by CMS \cite{Collaboration:2012tx} at this CL is $127 \leq m_h \leq
600 \, {\rm GeV}$. There is also a suggestive combination of
experimental hints in multiple final
state signatures pointing in both experiments  to a Higgs mass $m_h \simeq 124-126 \,  {\rm GeV}$. 
Although certainly not conclusive, the evidence for a light resonance
with this mass scale and approximately SM Higgs-like properties
has also been augmented by the
observation of 7 events (compared to an expected background of 2
events) within one ${\rm GeV}$ of 124 {GeV} in the $p p
\rightarrow h j j   \rightarrow \gamma \gamma jj$ channel at CMS
\cite{Collaboration:2012tw}, which would correspond to a SM Higgs boson
produced through vector boson fusion. We will use these experimental
results in this paper to determine to what degree current data is
selecting a SM Higgs boson or not.

These experimental hints, along with the lack of any clear evidence of
new states discovered to date at the LHC are suggestive that
an effective theory of the EWSB sector, including a light scalar
resonance and the approximate symmetries of $\rm SU(2)_c$ as well
as MFV, is currently an appropriate description of the data. We will
examine recent results from the LHC in this framework and ascertain up to
what degree experiment is selecting a SM Higgs doublet at present.  We perform a broad analysis of this question in Sections~II and~III
in terms of two parameters characterizing the deviation
from the SM value in the gauge and fermion couplings of a Higgs boson. We find two nearly degenerate minima
in the global fit.
The effective Lagrangian we employ emerges naturally in composite
Higgs scenarios. We comment on their interpretation in terms of current data.

The LHC results should be interpreted in the context of the indirect evidence for the SM
Higgs and its properties in EWPD. We examine the consistency of the results of our global fit to
LHC Higgs-like data and EWPD in Section IV. 

Finally, in Section V, we discuss how future measurements
can be presented in a manner that can more precisely and efficiently
refine the understanding of such an effective theory with the aim  of clarifying if the correct description of EWSB is the SM Higgs mechanism
or not. In particular we point out the utility of ratios of best fit
signal strengths which can be provided by the experimental
collaborations and which
have the ability to resolve the degeneracy in the two best fit regions we find
when examining LHC Higgs-like data. This is due to the fact that, although
an individual experimental signal can be faked by a Higgs-like degree of freedom,
it is difficult for such a degree of freedom to simultaneously reproduce the Higgs predictions in
other channels with different dependencies on effective couplings.

\section{The Effective Theory}
We consider an effective Lagrangian with a light scalar resonance, denoted as $h$
in the following, that includes the Goldstone bosons associated with
the breaking of $\rm SU(2) \times U(1)_Y \rightarrow U(1)_Q$ and the SM field content.
A minimal description of these degrees of freedom is given by an
effective chiral Lagrangian with a nonlinear realization of the $\rm
SU(2) \times U(1)_Y$ symmetry. The Goldstone bosons eaten by the $\rm
W^\pm, Z$ bosons are denoted by $\pi^a$ , where
$a = 1,2,3$, and are grouped as 
\bea
\Sigma(x) = e^{i \sigma_a \, \pi^a/v} \; ,
\eea
with $v = 246 \, {\rm GeV}$.  The $\Sigma(x)$ field transforms
linearly under $\rm SU(2)_L \times SU(2)_R$ as $\Sigma(x) \rightarrow
L \, \Sigma(x) \, R^\dagger$ where $L,R$ indicate the transformation
on the left and right under $\rm SU(2)_L$ and $\rm SU(2)_R$, respectively, while $\rm SU(2)_c$ is the diagonal subgroup of
$\rm SU(2)_L \times SU(2)_R$, under which the scalar resonance $h$
transforms as a singlet. A derivative expansion of such a theory
is given by \cite{Giudice:2007fh,Contino:2010mh,Grober:2010yv}
\bea
\mathcal{L} &=& \frac{1}{2} (\partial_\mu h)^2 - V(h) + \frac{v^2}{4} {\rm Tr} (D_\mu \Sigma^\dagger \, D^\mu \Sigma) \left[1 + 2 \, a \, \frac{h}{v} + b \, \frac{h^2}{v^2}  +  b_3 \, \frac{h^3}{v^3} + \cdots \right], \nn \\
&\,& - \frac{v}{\sqrt{2}} \, (\bar{u}_L^i \bar{d}_L^i) \, \Sigma \, \left[1 + c \, \frac{h}{v} +  c_2 \, \frac{h^2}{v^2}  + \cdots \right] 
\left(
\begin{array}{c} 
y_{ij}^u \, u_R^j \\ 
y_{ij}^d \, d_R^j 
\end{array} \right)  +  h.c., \nn  \quad \mbox{with} \quad\\
V(h) &=& \frac{1}{2} \, m_h^2 \, h^2 + \frac{d_3}{6} \, \left(\frac{3
    \, m_h^2}{v} \right) \, h^3 + \frac{d_4}{24} \left(\frac{3
    m_h^2}{v^2} \right) h^4 + \cdots \; . 
\label{eq:efflag}
\eea
Here we have adopted for later convenience the notation of composite
models \cite{Giudice:2007fh,Contino:2010mh,Grober:2010yv}. Minimal
Flavour Violation dictates that the sole source of flavour violation
are the Yukawa couplings and that $h$ is a singlet in
flavour space and that its couplings to fermions ($c,c_2,...$) are
flavour-universal. (Note that MFV is also compatible with $c_i$ proportional to a combination of the SM Yukawa matricies.)  We adopt this assumption, although we note that the LHC data is essentially only
sensitive to flavour violation linked to the large top coupling (scaled by $c$ from its SM value) at present.

The $\rm SU(2) \times U(1)_Y \rightarrow U(1)_Q$
subgroup of $\rm SU(2)_L \times SU(2)_R$ is weakly gauged in such theories.
We will fit the current data including the leading linear Higgs-like coupling effects in the
Lagrangian of  Eq.~(\ref{eq:efflag}). We will neglect, however, dimension five operators  $h \,G^{\mu \, \nu} G_{\mu \, \nu}$, 
$\,\,h \,W^{\mu \, \nu} W_{\mu \, \nu}$, $\,\, h \,B^{\mu \, \nu}
B_{\mu \, \nu}$ and other higher dimension operators in the fit\footnote{See Ref.~\cite{Manohar:2006gz} for a discussion on the impact of these operators on Higgs production.},
focusing on the effect of $a$ and $c$.
This can be justified by UV model building, for example in the composite Higgs case.  
In general, the coefficients $a,b,b_3,c,c_2,...$ are arbitrary numerical parameters subject to experimental constraints. Model building in the UV of this effective theory fixes relations between the parameters.

The completion of the theory with a SM Higgs boson fixes $a = b
= c = d_3 = d_4 =1$ and $b_3 = c_2 =0$, and higher order terms in the
polynomial expansion of the $h$ field vanish. In this case $h$ becomes part of a linear multiplet
\bea
U = \left(1 + \frac{h}{v} \right) \Sigma  \; ,
\eea
reducing the theory to the SM Higgs boson Lagrangian where the growth in
$\sqrt{s}$ of the high energy scattering cross section $W_L \, W_L
\rightarrow W_L \, W_L$ is exactly cancelled. For other choices of
these parameters the growth in $\sqrt{s}$ of high energy scattering
processes is only moderated. First we will explore to what degree
a light scalar resonance leading to the current experimental excesses
reported by CMS and ATLAS is consistent with a SM Higgs boson
interpretation or not.\footnote{Many proposals and discussions on establishing the
correctness of the SM Higgs hypothesis exist in the literature,
including Refs.~
\cite{Manohar:2006gz,Burgess:1999ha,Barger:2009me,Lafaye:2009vr,Bock:2010nz,Grinstein:2007iv,Goldberger:2007zk,
Bonnet:2011yx,Englert:2011aa,Duhrssen:2004cv,Campbell:2011iw}.}
Note that the fermiophobic Higgs hypothesis is frequently interpreted to correspond to the singular point
where $c = 0, a=1$, while a more general interpretation of this hypothesis is $c = 0$ with $a$ unfixed, but constrained by
current data. 
\section{Fitting Enhanced Cross Sections}
The experimental hints for the Higgs boson that have been reported to
date in each channel individually provide only marginal evidence.
The best fit values to the signal strengths $\mu = (\sigma \times {\rm Br})/(\sigma \times {\rm Br})_{SM}$ for
particular Higgs masses are summarized in Table I. 

In Table I we also give in the second column the expected mass sensitivity of the
various experimental signatures as well as in the third column the best
fit value for $\mu$ together with the $1 \sigma$ experimental
error and the $95 \%$ confidence level limit. If the mass resolution is not explicitly quoted, as in the $pp
\to Z Z^* \to l^+ l^- l^+ l^-$ and the $pp \to W W^* \to l^+ \nu l^-
\bar{\nu}$ results given by ATLAS, we display as an estimate the mass
sensitivity for these channels as quoted by CMS.  Since the mass
sensitivity is such that the best fit masses can reasonably be interpreted to overlap, we do not
introduce a correction factor in the fit to shift to a common mass
value. We also display, in column two, the local significance when
given by the experiments. In column four
we schematically list the leading sensitivity of the signal in terms
of the cross section rescaling and the effective branching ratios to
final states $XX$ given by $\rm Br_{XX} [a, c]$. The actual production
cross section rescalings are combinations including subdominant
channels, that can introduce further dependence on (a,c), see
Section~III.A. We do not include a look elsewhere effect as we are
performing a global fit for a particular mass value $m_h \approx 124$
GeV. The CMS photon measurements are split up in accordance with Ref.~\cite{Collaboration:2012tx}. The $\gamma\gamma$ events are classified by the conversion of the photon in the crystal - defining a parameter $R_9$ - and their location in the detector, being endcap - e or barrel - b. The data we fit to also have associated exclusion curves. We take these exclusions into account by another procedure described in the text. Note that the $\tau^+\tau^-$ searches at ATLAS are included in the exclusion analysis but not fit to in the signal strength best fit as the corresponding experimental error is not available.

The excesses of events with $\it approximately$ the same mass scale in
various channels are suggestive of a resonance, that could be
interpreted as evidence of a light Higgs boson. We will assume that
these excesses of events correspond to the same underlying physics and fit the data to 
discern the degree up to which the excesses are consistent with a SM
Higgs boson interpretation.\footnote{The local significance quoted in Table I should be taken as a
  reminder to the reader of the very marginal situation of the
  statistical significance of these signals at present.} Our procedure to perform a global fit to
the current data is as follows.
\begin{table} 
\setlength{\tabcolsep}{5pt}
\center
\begin{tabular}{c|c|c|c} 
\hline \hline 
Channel [Exp] &  $m_h [{\rm GeV}]$ (Local Significance) & $\mu$ ($\mu_L$)  & Scaling to SM
\\
\hline
$pp \rightarrow \gamma \, \gamma \,\, [{\rm ATLAS}]$  & $126.5 \pm 0.7 \, \, \, (2.8\, \sigma)$ \cite{Collaboration:2012sk} & $2^{+0.9}_{-0.7}$ \cite{atlas1} (2.6)& $ \sim c^2  \, {\rm Br}_{\gamma \, \gamma}[a,c]$ 
\\
$pp \rightarrow  Z \, Z^\star \rightarrow \ell^+ \, \ell^- \, \ell^+ \, \ell^- \,\, [{\rm ATLAS}] $ & $126 \pm \sim 2 \%  \, \, \, (2.1\, \sigma)$  \cite{Collaboration:2012sk} & $1.2^{+1.2}_{-0.8}$ \cite{atlas1} (4.9) & $ \sim c^2  \, {\rm Br}_{ZZ}[a,c]$ 
\\
$pp \rightarrow  W \, W^\star \rightarrow \ell^+ \, \nu  \, \ell^- \, \bar{\nu} \,\, [{\rm ATLAS}] $ & $126\pm \sim 20 \% \, \, \, (1.4 \, \sigma)$  \cite{Collaboration:2012sk} & $1.2^{+0.8}_{-0.8}$ \cite{atlas1} (3.4) & $ \sim c^2  \, {\rm Br}_{WW}[a,c]$ 
\\
$pp \rightarrow  \gamma \, \gamma  \, jj \,\, [{\rm CMS}]$  & $124 \pm 3 \%$ \cite{Collaboration:2012tx,Collaboration:2012tw} & $3.7^{+2.5}_{-1.8}$ \cite{Collaboration:2012tw}& $ \sim a^2  \, {\rm Br}_{\gamma \gamma}[a,c]$ 
\\
$pp \rightarrow  \gamma \, \gamma  [{\rm CMS,b},R^{\rm min}_9>0.94]$ & $124 \pm 3 \%$ \cite{Collaboration:2012tx,Collaboration:2012tw} & $1.5^{+1.1}_{-1.0}$ \cite{Collaboration:2012tw} & $ \sim c^2  \, {\rm Br}_{\gamma \, \gamma}[a,c]$  
\\
$pp \rightarrow  \gamma \, \gamma  [{\rm CMS,b},R^{\rm min}_9<0.94]$ & $124 \pm 3 \%$ \cite{Collaboration:2012tx,Collaboration:2012tw} & $2.1^{+1.5}_{-1.4}$ \cite{Collaboration:2012tw} & $ \sim c^2  \, {\rm Br}_{\gamma \, \gamma}[a,c]$ 
\\
$pp \rightarrow  \gamma \, \gamma  [{\rm CMS,e},R^{\rm min}_9>0.94]$ & $124 \pm 3 \%$ \cite{Collaboration:2012tx,Collaboration:2012tw} & $0.0^{+2.9}$ \cite{Collaboration:2012tw} & $ \sim c^2  \, {\rm Br}_{\gamma \, \gamma}[a,c]$ 
\\
$pp \rightarrow  \gamma \, \gamma  [{\rm CMS,e},R^{\rm min}_9<0.94]$ & $124 \pm 3 \%$ \cite{Collaboration:2012tx,Collaboration:2012tw} & $4.1^{+4.6}_{-4.1}$ \cite{Collaboration:2012tw} & $ \sim c^2  \, {\rm Br}_{\gamma \, \gamma}[a,c]$ 
\\
$pp \rightarrow  Z \, Z^\star \rightarrow \ell^+ \, \ell^- \, \ell^+ \, \ell^- \,\, [{\rm CMS}] $ & $126 \pm 2 \% \,\, \, (1.5 \, \sigma)$ \cite{cms1,Collaboration:2012tw}  & $0.5^{+1.0}_{-0.7}$ \cite{Collaboration:2012tx} (2.7) & $ \sim c^2  \, {\rm Br}_{ZZ}[a,c]$
\\
$pp \rightarrow  W \, W^\star \rightarrow \ell^+ \, \nu  \, \ell^- \, \bar{\nu} \,\, [{\rm CMS}] $ & $126 \pm 20 \%$ \cite{Collaboration:2012tx,Collaboration:2012ty} & $0.7^{+0.4}_{-0.6}$ \cite{Collaboration:2012tx} (1.8) & $ \sim c^2  \, {\rm Br}_{WW}[a,c]$  \\
$pp \rightarrow  b \, \bar{b} \,\, [{\rm CMS}] $ &   $124  \pm 10 \%$ \cite{Collaboration:2012tx}  & $1.2^{+1.4}_{-1.7}$ \cite{Collaboration:2012tx} (4.1) & $ \sim a^2  \, {\rm Br}_{b\bar{b}}[a,c]$ 
\\
$pp \rightarrow  \tau \, \bar{\tau} \,\, [{\rm CMS}]$ &  $124  \pm 20 \%$\cite{Collaboration:2012tx}   & $0.8^{+1.2}_{-1.7}$ \cite{Collaboration:2012tx} (3.3)& $ \sim c^2  \, {\rm Br}_{\tau \bar{\tau}}[a,c]$ 
\\
\hline \hline
\end{tabular}
\caption{Summary table of reported signatures with events related to the Higgs mass scale of interest ($m_h\simeq 124$ GeV) where excess events have been reported.}
\label{table:tcorrections} \vspace{-0.35cm}
\end{table}

First we fit to reported values of $\mu^i$ including the
deviations in the SM predictions by allowing the parameters $a$ and
$c$ to deviate from their SM values of 1. We include in the fits the
effects of modified production cross sections and branching ratios due
to the rescaling of the SM couplings by the parameters $a$ and $c$.  
In order to carry out these fits we are required to make a set of
assumptions, which are summarized and discussed in
the Appendix. We will illustrate the sensitivity of the fit
to the various assumptions by varying them in the results presented in Section \ref{resultfit}. 

For example, consider the case of the event yield used to construct each
$\mu^i$ for $pp \rightarrow \gamma \, \gamma$, which
in this discussion we
will assume is only produced through gluon fusion. The SM prediction for the Higgs
boson producing such events with an integrated Luminosity $\int
\mathcal{L} \,  dt$  can be schematically written as
\bea
\int_{\rm cuts} \mathcal{L} \,  dt \, \, \sigma_{gg\rightarrow h} \,
{\rm Br}(h \rightarrow \gamma \gamma) \; . 
\eea
We will assume that the effects of the coefficients $a,c$ are to
simply rescale the number of events in the various signal cross
sections. We neglect small shape differences in the differential
distributions that could affect the event yield as $(a,c)$ deviate
from $(1,1)$. Then the integration over luminosity with phase space
cuts will essentially cancel in the constructed theoretical prediction
for the ratio $\mu^i$. The sole effect of the deviation
from $(1,1)$ for $(a,c)$ will be to generate an excess/suppression of
events compared to the SM prediction. This can be directly fit to the
reported experimental best fit value for this ratio. This is the
procedure we will adopt. 

We construct a $\chi^2$ measure for a two parameter fit in the following way. We define the matrix $\C$ as the covariance matrix of the
observables, and $\Delta \, \theta_i$ as a vector of the difference
in the observed and predicted value of the ratio, as a function of
$(a,c)$. The $\chi^2$ measure is then given by 
\bea
\chi^2 = (\Delta \theta_i)^T  \, (\C^{-1})_{i,j} \, (\Delta \theta_j)
\; .
\eea
The minimum $\chi^2_{min}$ is determined, and the $65 \%, 90 \%$ and
$99 \%$ best fit CL regions are given by $\Delta \chi^2 < 2.1,
\,\,4.61, \, \, 9.21$, respectively, for $\chi^2 = \chi^2_{min} + \Delta \chi^2$.
The confidence level regions are defined by the cumulative distribution
function for a two parameter fit. The matrix $\C$ is taken to be diagonal with the square of
the $1 \, \sigma$ theory and experimental errors added in quadrature 
for each observable in the diagonal element. As correlation
coefficients are currently not supplied by the experimental
collaborations, off-diagonal correlation coefficients are
neglected. For the experimental errors we 
use the quoted $1 \, \sigma$ errors on the reported signal
strength. The errors $\delta \mu^i$ for the individual channels $i$
are made symmetric by taking $\delta \mu^i= \sqrt{(\delta \mu^{i+})^2 +
  (\delta  \mu^{i-})^2}/\sqrt{2}$, where $\mu^i = \sigma^i/\sigma^i_{SM}$. For
theory predictions of the cross section values and related errors, 
we use the numbers given on the webpage of the LHC
Higgs Cross Section Working Group \cite{Dittmaier:2011ti} 
for $m_h = 124 \,{\rm GeV}$ and $\sqrt{s}=7$~TeV.\footnote{The theory errors are currently subdominant compared to the large experimental uncertainties.} We symmetrize the total cross
section error and propagate the error to get a theory error on
$\mu$. Taking  
\bea
\mu = \frac{\sum_i \, r^i(a,c) \, \sigma^i \, \times {\rm Br_i}(a,c)}{\sum_i \, \sigma^i \,  \times {\rm Br_i}_{SM}} \; ,
\eea
with $r^i(a,c)$ the appropriate rescaling factor for each cross
section and defining the error on
each $\sigma^i$ to be $\delta \sigma^i$, we determine the error $\delta \mu$ by
\bea
\delta \mu =\mu \sqrt{\left[\frac{\sqrt{\sum_i \, (r^i)^2 \,
        (\delta \sigma^i)^2}}{\sum_i \, r^i \, \sigma^i} - \frac{\sqrt{\sum_i
        \, (\delta \sigma^i)^2}}{\sum_i \, \sigma^i} \right]^2} \; .
\eea

For each search channel, ATLAS and CMS provide
\cite{Collaboration:2012sk,Collaboration:2012tx,Collaboration:2012tw,Collaboration:2012ty}
an exclusion upper limit $\mu_L^i$ on each `signal strength' so that $\mu^i>\mu_L^i$ is excluded at
95\% C.L. This value is reported in Table I. The final exclusion limit $\mu_L$ from combining all
channels is also provided as a function of the Higgs mass, with a SM
Higgs boson mass being excluded whenever $\mu_L<1$. We obtain such combined limits
by a simple $\chi^2$ procedure, solving for $\mu_L$ the equation
\bea\label{applim}
\sum_i\frac{(\mu_L-\hat\mu)^2}{(\mu_L^i-\hat\mu)^2}-
\sum_i\frac{\hat\mu^2}{(\mu_L^i-\hat\mu)^2}=1\; ,
\eea
where $\hat\mu$ is the average of the individual $\hat{\mu}^i$, the measured signal strengths for each channel
(which make $\mu_L^i$ larger when they are nonzero)\footnote{We contrast this semi-empirical 
approximate formula
with a more precise determination of the combined limit in Appendix~B.}. When there are
no excesses (or for the purpose of calculating expected limits) one
simply sets $\hat\mu^i=0$, in which case our simple recipe combines in quadrature the limits
from different channels. Again we neglect correlations in the measured
limits on the signal strengths. In applying this procedure to our
case, note that we have mapped the reported $\rm CL_s$ exclusion
curve as 
\bea
\mu^i_{SM} \rightarrow \mu^i(a,c)
\eea
through rescaling the production cross sections and the decay
widths. The exclusion regions are based on all released Higgs search
analysis channels from the experiments given 
in December 2011 and updated in February 2012, and are comprised of the signals listed in Table~I (plus the ATLAS $h\rightarrow \tau^+\tau^-$ analysis). The results of the fit are dependent on the SM values used for the
masses of the known particles, gauge coupling constants in the SM,
etc. We summarize the SM inputs, which we have used,  in
Appendix A.

\subsection{Results} \label{resultfit}

\begin{figure}[tb]
\includegraphics[width=0.45\textwidth]{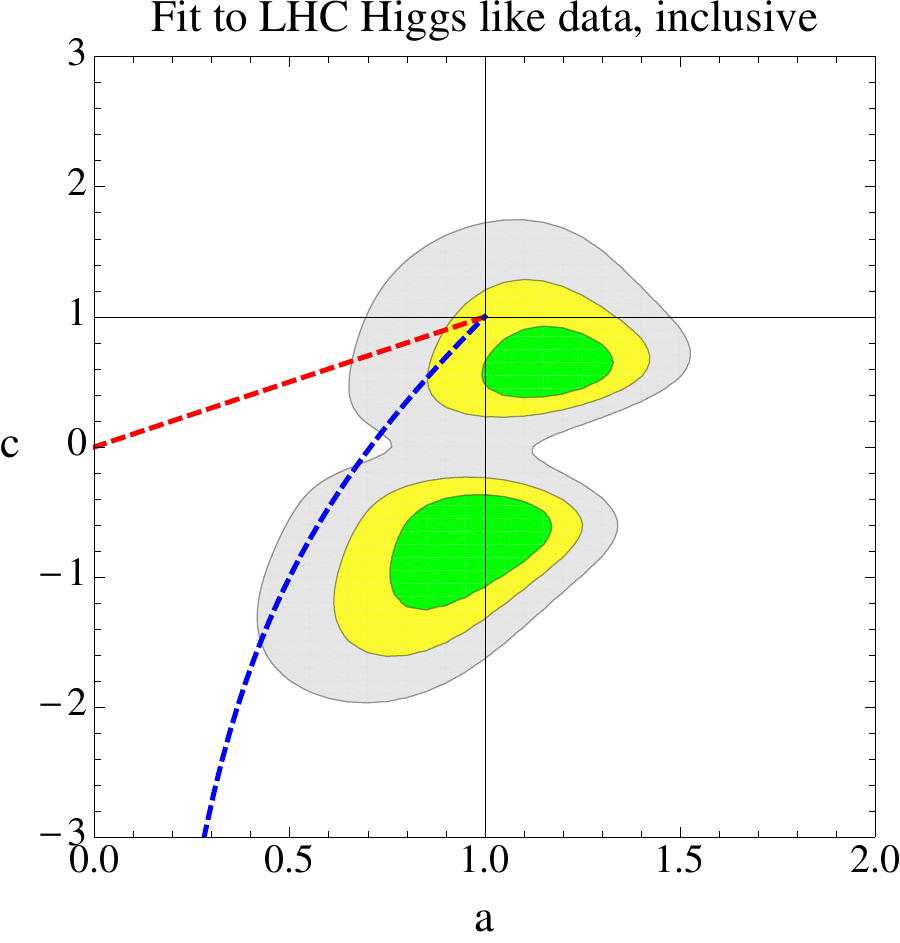}
\includegraphics[width=0.45\textwidth]{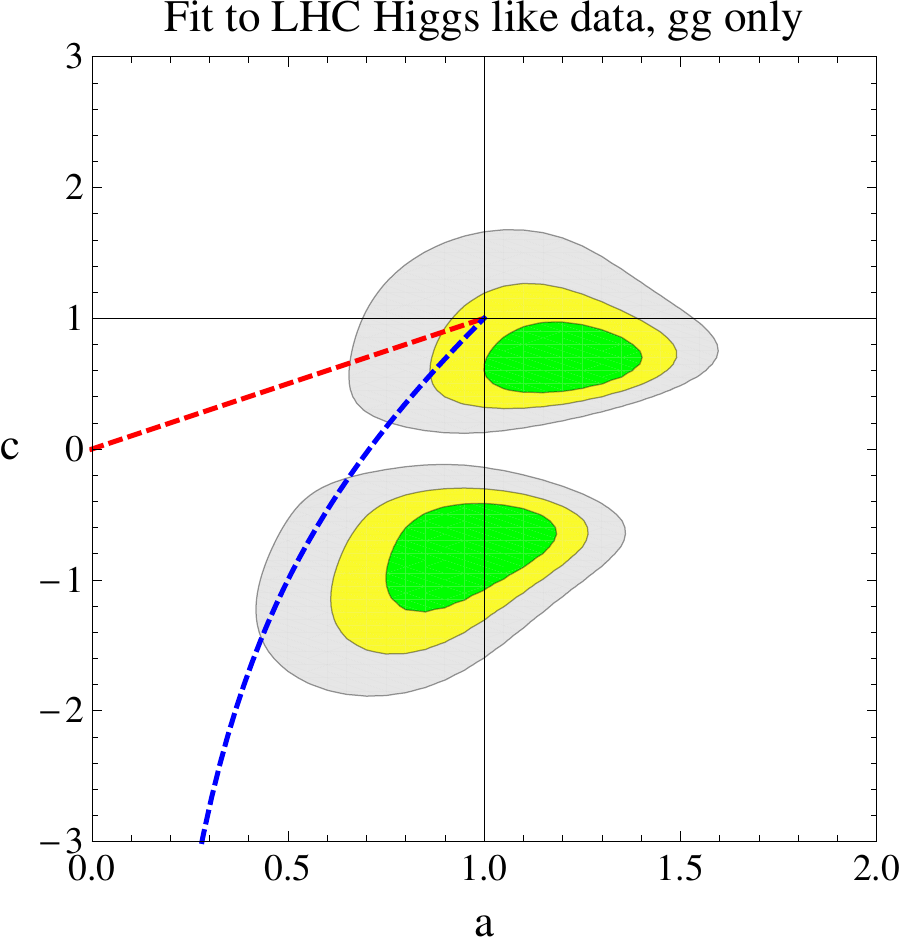}
\caption{Global fit results in the $(a,c)$ plane for all best fit
  $\sigma/\sigma_{SM}$ values given by ATLAS and CMS. The SM Higgs
  boson with a linear realization of the EW chiral Lagrangian
  corresponds to the point $(1,1)$ and is within the yellow $90 \%$ CL
  region. The $65 \%$ and $99\%$  CL regions correspond to the green and
  light gray shaded regions. The plot on the left includes all
  production channels, the plot on the right only includes the
$gg$ production channel in the signals with inclusive Higgs production, rescaled as described in the text. The
$\chi^2/{\rm d.o.f} \simeq 0.99$ for the left plot, while $\chi^2/{\rm
  d.o.f} \simeq 1.03$ for the right plot. We also show two lines characterizing the relationship between the parameters a,c
  in the minimal composite Higgs scenarios of Refs.~\cite{Agashe:2004rs,Contino:2006qr},
  red dashed (for MCHM4) and blue dashed (for MCHM5). In Eq.~(\ref{eq:efflag}), rephasing of the Higgs field $h \rightarrow - h$
  maps the best fit regions shown to other physically equivalent regions in the $(a,c)$ space.}\label{Fig.global} 
\end{figure}

Using the above described procedure we perform a global fit to the currently available data, with results 
shown in Fig.~\ref{Fig.global}. The global fit results are combined
with the exclusion contours in Fig.~\ref{Fig.globalexc}. The current
data is consistent with a SM Higgs boson interpretation which lies on the
$82 \%$ CL contour. The global quality of the fit
is $\chi^2/{\rm d.o.f} \approx 1$. There are two best fit regions.
For each signal strength to which we fit, there is a mapping from the SM
point in $(a,c)$ space, given by $(1,1)$, to a nearly degenerate point in terms of $\Delta \chi^2$
with differing values of $(a,c)$. This mapping can be understood as following from the contours of constant production shown in Fig.~\ref{Fig.signalcontour} which show an approximate $c \leftrightarrow -c$ symmetry in the
production of the Higgs and its subsequent branching ratios. This symmetry is broken by the interference of 
loop diagrams in various channels, with the strongest breaking of this approximate symmetry
in the $\gamma \, \gamma$ final state. The best fit values for the two best fit regions in the inclusive plot  are $(0.99,-0.66,1.99)$ and $(1.2,0.60,2.6)$
in terms of $(a,c,\chi^2_{min})$. The SM point for comparison is $(1,1,5.44)$.
It is of interest to determine a means by which the best fit region degeneracy can be resolved.
We discuss such an approach in Section V.

Note that in  Figs.~\ref{Fig.global}, \ref{Fig.globalexc}
we present results which we label as `inclusive' or `$gg$ only'. These fits differ in how the production cross sections
are treated. For `$gg$ only' in the signals with $\gamma \gamma$, $\tau^+ \,
\tau^-$ final states, we only rescale the dominant gluon fusion production channel, neglecting subdominant
channels. While for results labeled as inclusive the following production channels are included for each signal:

\begin{figure}[tb]
\includegraphics[width=0.45\textwidth]{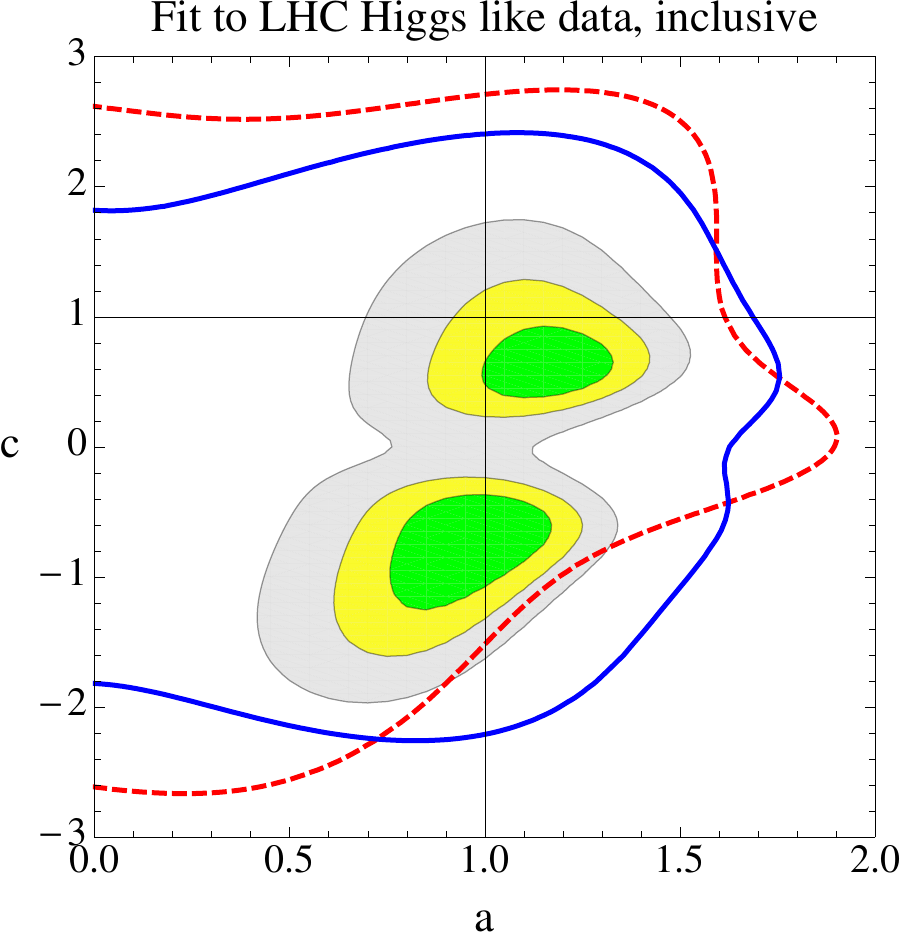}
\caption{Global fit results in the $(a,c)$ plane for all best
  fit $\sigma/\sigma_{SM}$ values given by ATLAS and CMS, taking into
  account all production channels. 
Also shown are the exclusion contours in the $(a,c)$ plane
determined by mapping the SM exclusion into an effective exclusion of
$(\sigma/\sigma_{SM})[a,c] <1$. 
Again the $65,90,99 \%$ CL regions correspond to the green, yellow and
gray regions in the plots. The exclusion curve derived from ATLAS data
is given by the red (dashed) line (with the region to the right of the line excluded), the
CMS exclusion curve is the solid blue line.}\label{Fig.globalexc}
\end{figure}
\begin{itemize}
\item[(i)] $\rm W^+ \, W^-/ZZ$ production via gluon fusion and vector boson fusion,
\item[(ii)] $\gamma \gamma$ production via gluon fusion, vector boson fusion (VBF), $\rm t \bar{t} h$ production, and associated production with $\rm W^\pm$ and $\rm Z$.
(Note that the $\gamma \gamma$ events that have associated jets, interpreted to come from VFB, are treated exclusively.)
\item[(iii)] $\rm b \bar{b}$ production is summed over associated $\rm h \, W^\pm$ and $\rm h \, Z$ production,
\item[(iv)] $\tau^+ \, \tau^-$ production is summed over gluon fusion,
  vector boson fusion, $\rm t \bar{t} h$ production, and associated production with $\rm W^\pm$ and $\rm Z$.
Previous analyses at CMS are reported to only include the VBF initial
state as the tagging jets are used to eliminate Drell--Yan $\rm Z
\rightarrow  \tau^+ \, \tau^-$ events. The updated results
use a modelling of the Drell--Yan
spectrum based on measurements of Drell--Yan produced $\rm Z
\rightarrow  \mu^+ \, \mu^-$ events, so that the updated analysis is more inclusive.
\end{itemize}

We do not find dramatic differences in the fits using the two different procedures,
and thus are lead to consider our approximations used in performing the rescalings
to be satisfactory. See Appendix A for a detailed discussion.

\begin{figure}[tb]
\includegraphics[width=0.33\textwidth]{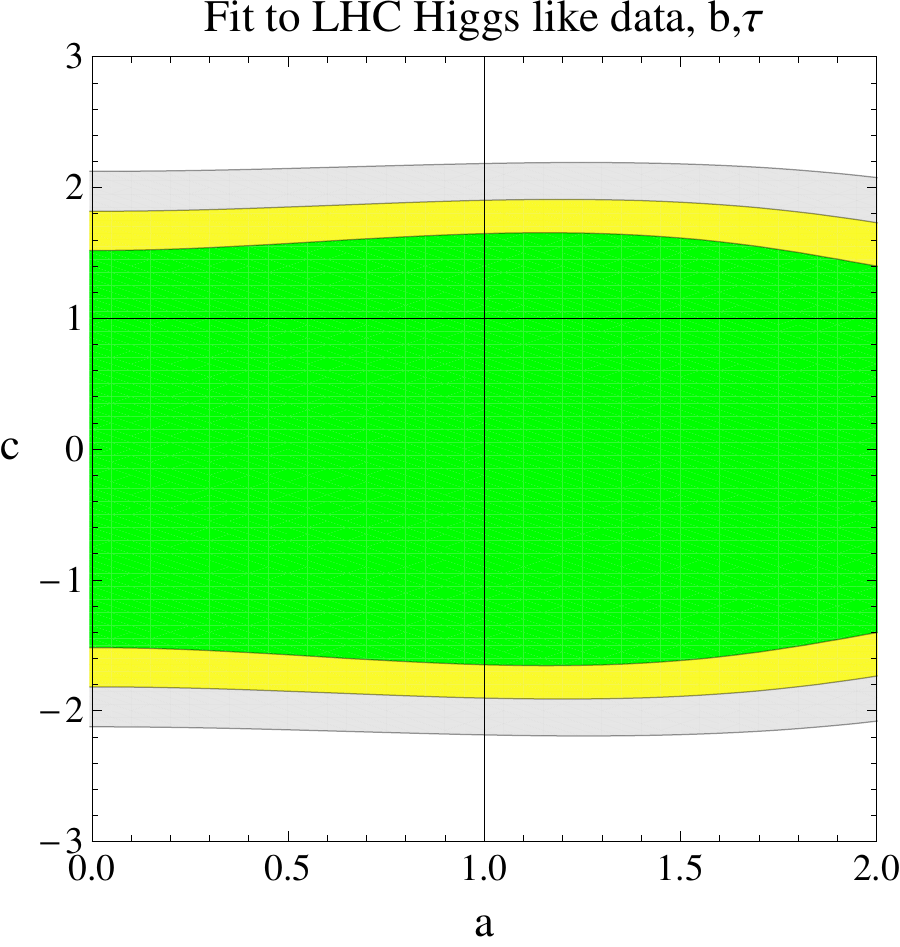}
\includegraphics[width=0.33\textwidth]{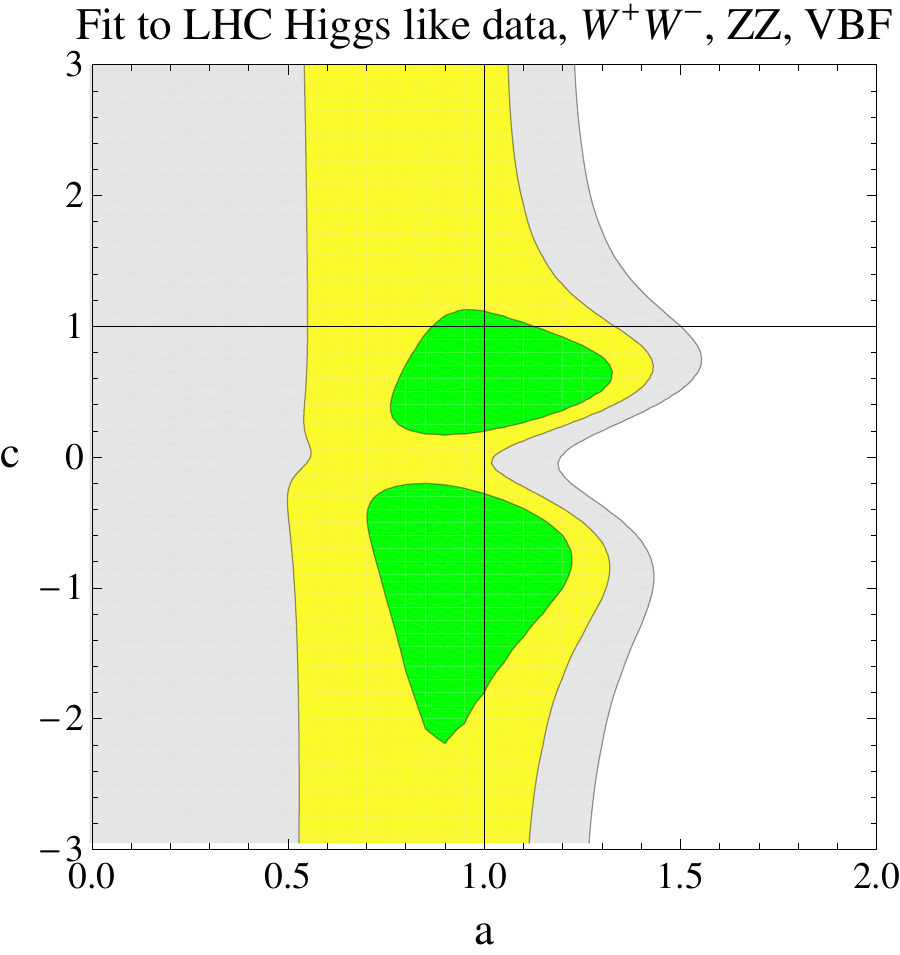}
\includegraphics[width=0.33\textwidth]{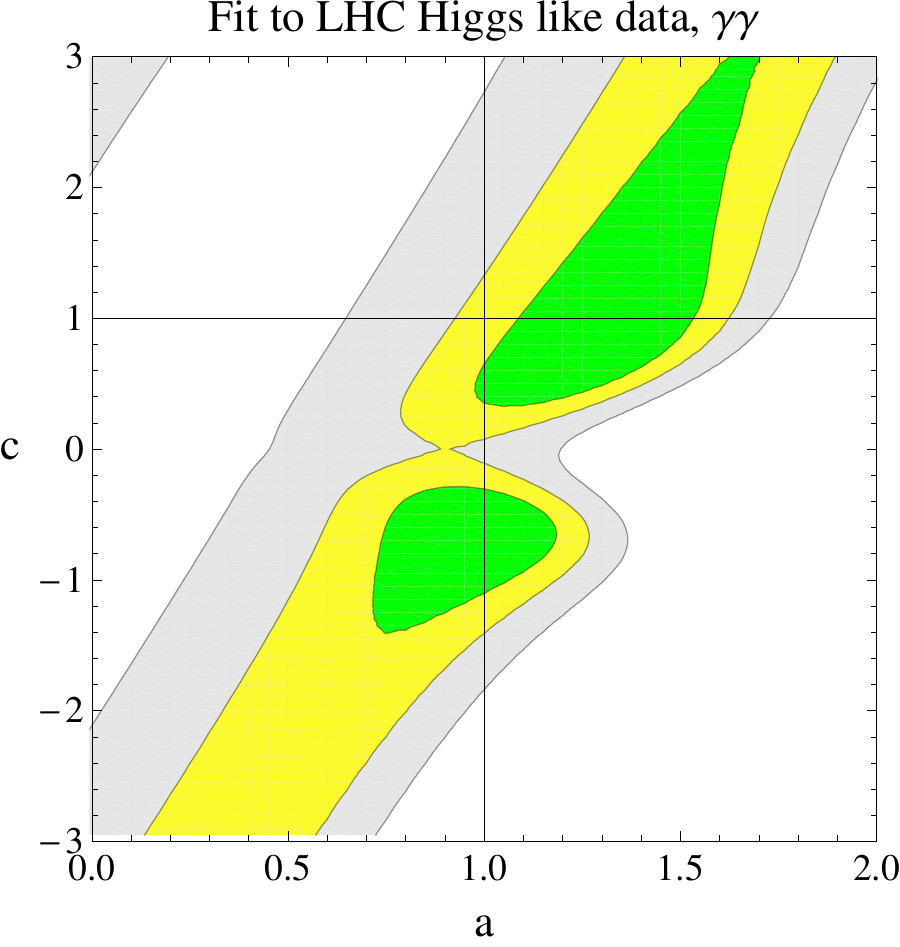}
\caption{Global fit results in the $(a,c)$ plane for all reported best
  fit $\sigma/\sigma_{SM}$ values given by ATLAS and CMS using subsets
  of the data. The conventions for the plots are the same as in
    the previous figures.
Fitting the data in subsets degrades the fit. The left figure testing the $b \, \bar{b}$ and $\tau^+ \, \tau^-$ data essentially is unconstrained with $\chi^2/{\rm d.o.f} \simeq 0$,
the middle plot with the combination of $\rm W^+ \, W^-$, $\rm Z \, Z$ and vector boson fusion based $\gamma \gamma$ has a $\chi^2/{\rm d.o.f} \simeq 0.28$ while the right plot based on all $\gamma \gamma$ excesses
has a $\chi^2/{\rm d.o.f} \simeq 0.62$.}\label{Fig.subsets}
\end{figure}

As can be inferred from Fig.~\ref{Fig.global} and Fig.~\ref{Fig.globalexc}, the
direct fit to the data and the exclusion curves are selecting the
same region of parameter space. We also performed exclusive fits to
the following subsets of data. We combined the $b \ \bar{b}$ and $\tau^+ \tau^-$ 
data for a test of the fermion couplings and combined the 
$\rm W^+ \, W^-$, $\rm ZZ$ data as well as the vector boson fusion based $\gamma
\gamma$ data for a test of anomalous gauge couplings. Furthermore, we
did a joint fit to the $\gamma \gamma$ best fit signal
strengths. The results are
shown in Fig.~\ref{Fig.subsets}.  As can be inferred from the figures,
the data are insensitive to the fermion and gauge couplings in the
$\tau \, \tau$ and $b \, \bar{b}$ channels, the sensitivity in the combination of the $\rm W^+ \, W^-$, $\rm ZZ$ data as well as the vector boson fusion based $\gamma
\gamma$ data is more significant, with a -roughly- $c$ symmetric region being selected and large enhancements of
the gauge boson couplings being disfavoured. The results of the $\gamma \gamma$ fits are most interesting as
the interplay of the top and gauge boson loops in the $\gamma \gamma$ decay amplitude select a skewed region in the $(a,c)$
space. One can also see that the $\gamma \gamma$ data are most constraining at present. It cannot be overemphasized that
the signal significances being fit to globally, and particularly when fitting to subsets of the data, are marginal.
Strong conclusions are premature and the best fit values for $\chi^2/{\rm d.o.f}$ in fitting to subsets of the data are poor. 

\section{Incorporating EWPD}
Deviations in the properties of a light scalar resonance from the
properties of the SM Higgs boson also affect other
low energy precision tests, in particular EWPD,  which are
sensitive to loop corrections to the gauge boson propagators due to
Higgs boson loops as well as the
SM gauge bosons loops. If the properties of a scalar deviate from the properties of the SM Higgs in the coupling to the SM gauge bosons,
the high energy behavior of the theory is modified, and the decoupling
of the longitudinal mode of the EW gauge bosons is no longer
exact. 

This effect manifests itself through the high energy behavior 
of the longitudinal degrees of freedom in high energy scattering still
growing with energy, and in EWPD through loop diagrams involving the
longitudinal degrees of freedom of the gauge bosons
\cite{Barbieri:2007bh}. In EWPD, the corrections to the gauge boson propagators
can be expressed in terms of shifts of the parameters STU \cite{Holdom:1990tc,Peskin:1991sw,Altarelli:1990zd} given by
\bea
\Delta S &=& \frac{-(1 - a^2)}{6 \, \pi} \, \log \left(\frac{m_h}{\Lambda}\right),  \quad \quad  \Delta T = \frac{3(1 - a^2)}{8 \, \pi \, \cos^2 \theta_W} \, \log \left(\frac{m_h}{\Lambda}\right),  \quad \quad  \Delta U = 0.
\eea

\begin{figure}[htb]
\includegraphics[width=0.45\textwidth]{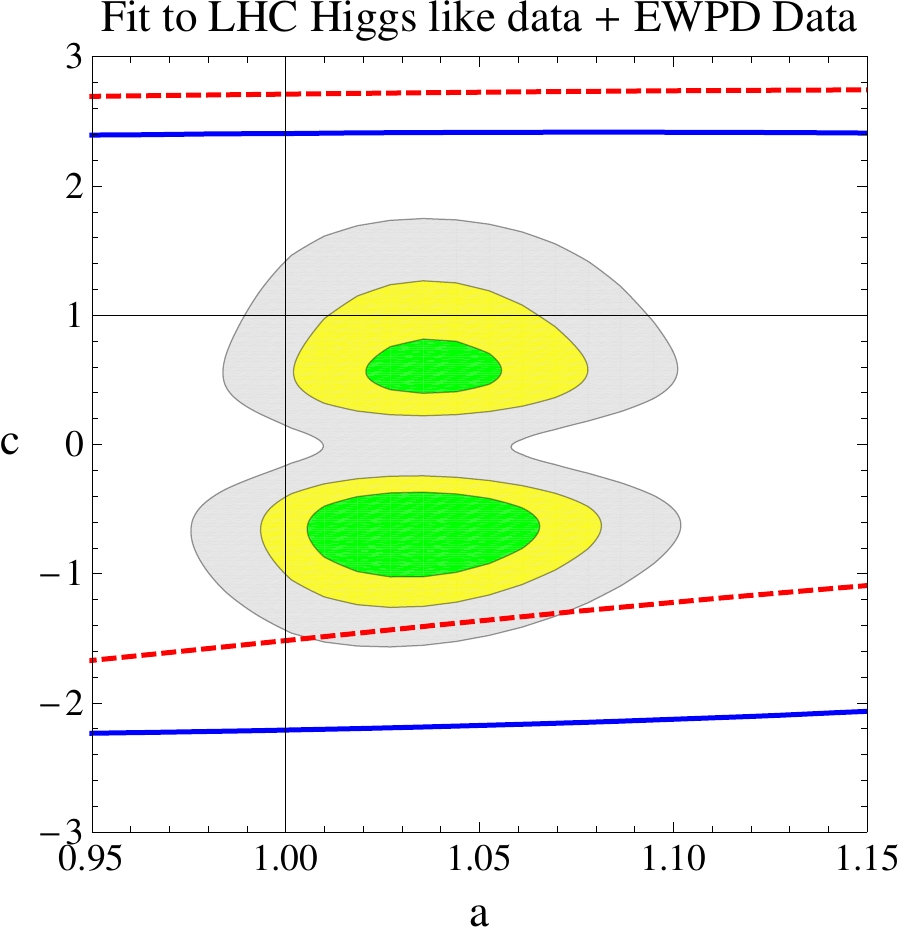}
\caption{Global fit results in the $(a,c)$ plane for all best
  fit $\sigma/\sigma_{SM}$ values given by ATLAS and CMS
  combined with EWPD. The plot follows the same convention as the previous plots.
The $\chi^2/{\rm d.o.f} = 1.3$. As before the blue  (red) curve is the exclusion region from CMS (ATLAS). Again
the shown exclusion curves  are determined from inclusive production.}\label{Fig.EW}
\end{figure}

Here we have introduced a cutoff scale $\Lambda$, which
approximately represents the mass of the new states that are required in this
framework to $\it unitarize$ longitudinal gauge boson scattering at a scale given by $\Lambda \sim 4 \, \pi \, v/\sqrt{|1 - a^2|}$. For EWPD we use the 
results of the ${\it Gfitter}$ collaboration \cite{Baak:2011ze}
\bea
S = 0.02 \pm 0.11,  \quad \quad  T = 0.05 \pm 0.12,  \quad \quad U =
0.07 \pm 0.12 \;.
\eea
And the correlation coefficient matrix is given by
\bea
C = \left(
\begin{array}{ccc} 
1 & 0.879 & -0.469 \\ 
0.879 & 1 & -0.716 \\
 -0.469 & -0.716 & 1 \\
\end{array} \right).
\eea
We perform joint fits to LHC data and EWPD by adding the corresponding entries to an enlarged covariance matrix (including STU in terms of $a$) in the global
fit. Correlations among EWPD observables are included. We then perform a new global minimization and joint fit. The results are given in Fig \ref{Fig.EW}.
We have included a correction to $\rm STU$ of the form $(\Delta  S, \Delta T, \Delta U) = (0.003,-0.002,-0.0001)$ due to shifting the best fit value of the Higgs mass in these
results from $120$~GeV to $124 \, {\rm GeV}$ using the exact
expression for the one-loop Higgs boson contribution to $\rm STU$.
When a joint fit to current LHC data is combined with EWPD under the
assumption that $m_h \sim 124 \, {\rm GeV}$ and with no other states in the EWPD
fit, the tension in the fit slightly increases. The $\chi^2/{\rm d.o.f}$ rises to 1.3, with EWPD favouring values of $a>1$ as the STU parameters do 
not have vanishing central values for $m_h= 124 \, {\rm GeV}$. The allowed regions are consistent with
the exclusion curves as shown in Fig \ref{Fig.EW}.
\section{Discussion}

In examining possible signal data for the Higgs boson, such as
the reported best fit values of $\sigma/\sigma_{SM}$ from each
experimental collaboration, the first task should be to clarify if
a resonance is consistent with a SM Higgs interpretation or not. From an effective field theory perspective, there is no particular motivation for the couplings of the Higgs to be identical to one predicted by the SM. 
The hierarchy problem strongly suggests the existence of new states with a relatively low mass, coupled to the Higgs. Integrating out these states, one expects the properties of the Higgs will change.
Nevertheless, when approaching the data in effective field theory, the symmetries
that are known to be at least approximately present at the weak scale
are already highly constraining,
as discussed in Section II. One is therefore lead to the effective description of the data which we have adopted\footnote{The robustness of the requirement of exact MFV is less significant than $\rm SU(2)_c$ in the EWSB sector.}, and global fits to precision Higgs data will allow the SM
hypothesis of EWSB to be experimentally tested in the near future.  
\begin{figure}[tb]
\includegraphics[width=0.45\textwidth]{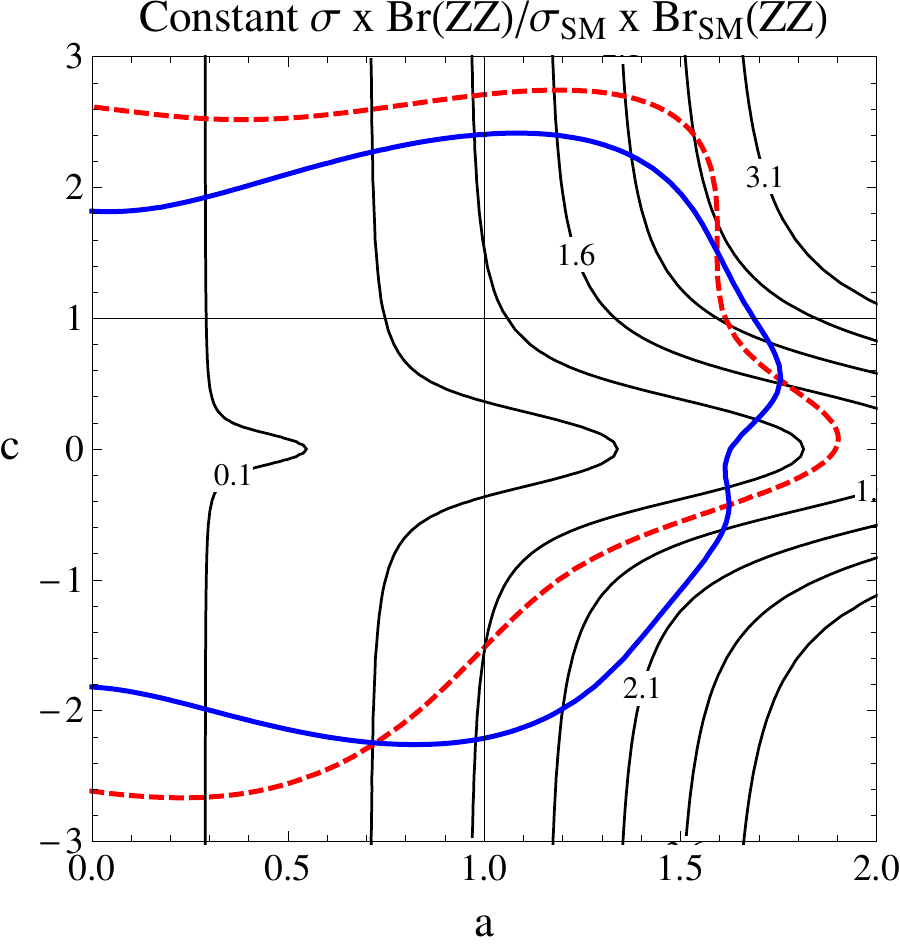}
\includegraphics[width=0.45\textwidth]{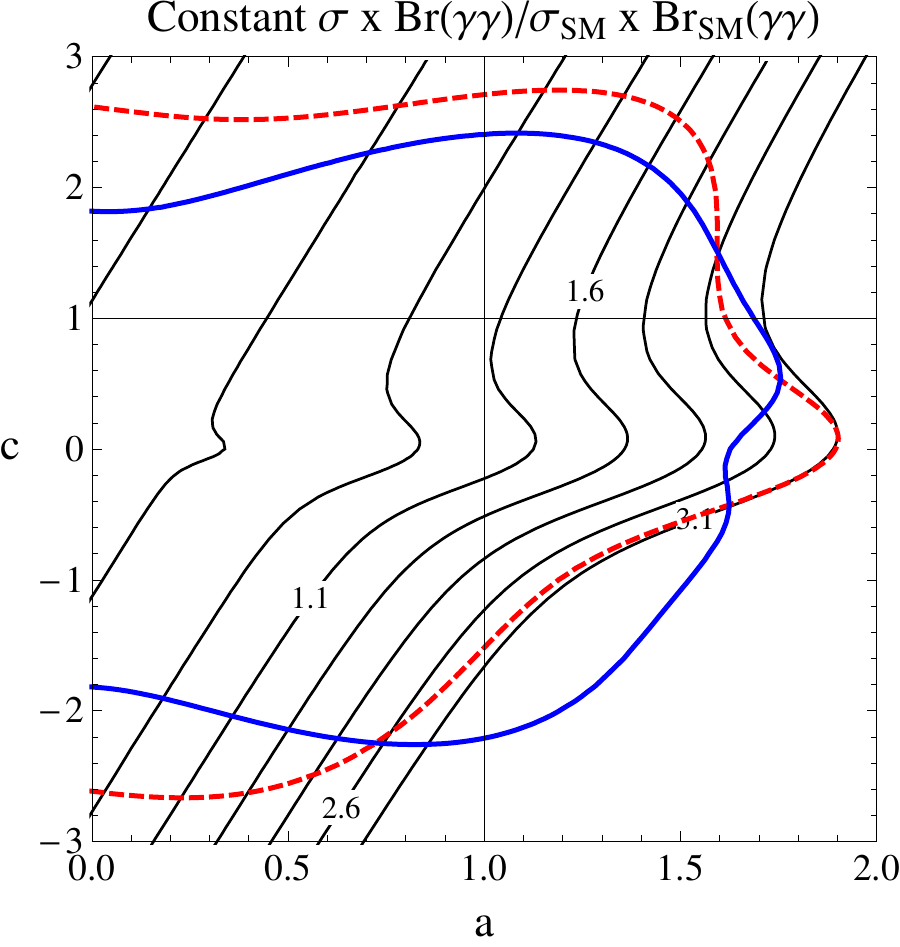}
\includegraphics[width=0.45\textwidth]{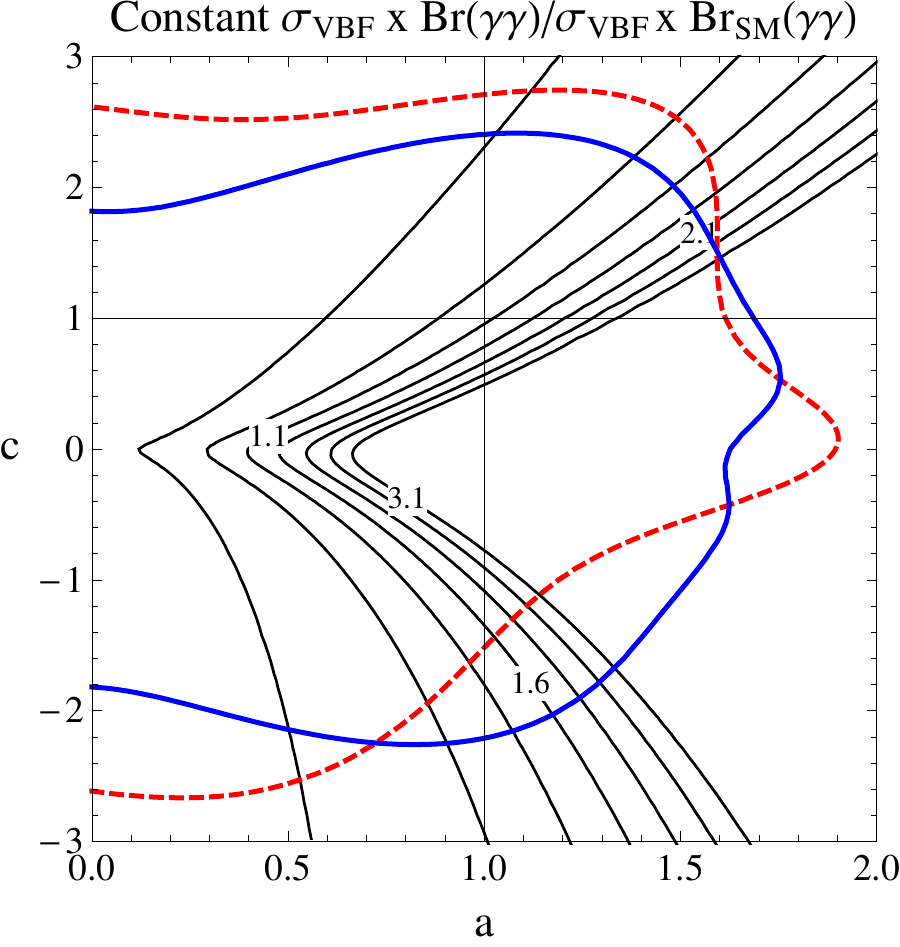}
\includegraphics[width=0.45\textwidth]{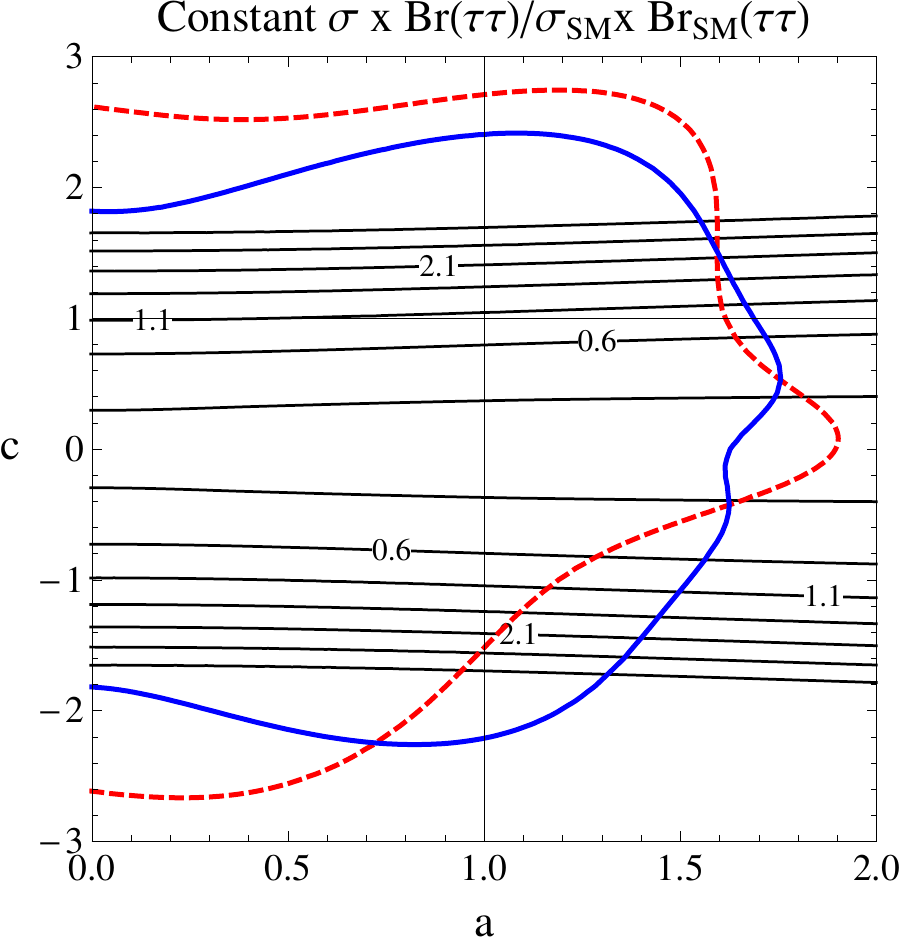}
\caption{Contours of constant signal production in the $(a,c)$
  plane. We have also included the exclusion curves which we
  have derived from ATLAS and CMS data.}\label{Fig.signalcontour}
\end{figure}

As the LHC will be running at $8 \, \, {\rm TeV}$, the contours of
constant $\mu^i =[\sigma\times {\rm BR}](h \rightarrow XX) [a,c]/[\sigma\times {\rm
  BR}]_{SM}(h \rightarrow XX)$ in a final state $X$
  are constructed in the $(a,c)$ plane
for this centre of mass (c.m.) energy in Fig.~\ref{Fig.signalcontour}. We
restrict ourselves to a subset of the channels used in this analysis. 
The prospects of reducing the degeneracy in the best fit regions 
relies mostly on further
observations related to signal events in $\sigma \times {\rm BR}(h
\rightarrow \gamma \gamma)$. 
Here we have rescaled using the previously defined `inclusive' rescaling and combined the production channels that contribute to the analyses
of signal channels. Note that each individual channel has a
mapping away from the SM point of $(1,1)$ in the $(a,c)$ plane to a family of degenerate points.  This leads to two best fit areas,
  which have to be resolved. This can be done by realizing that the mapping is not the
  same when comparing among different channels. 
\begin{figure}[tb]
\includegraphics[width=0.49\textwidth]{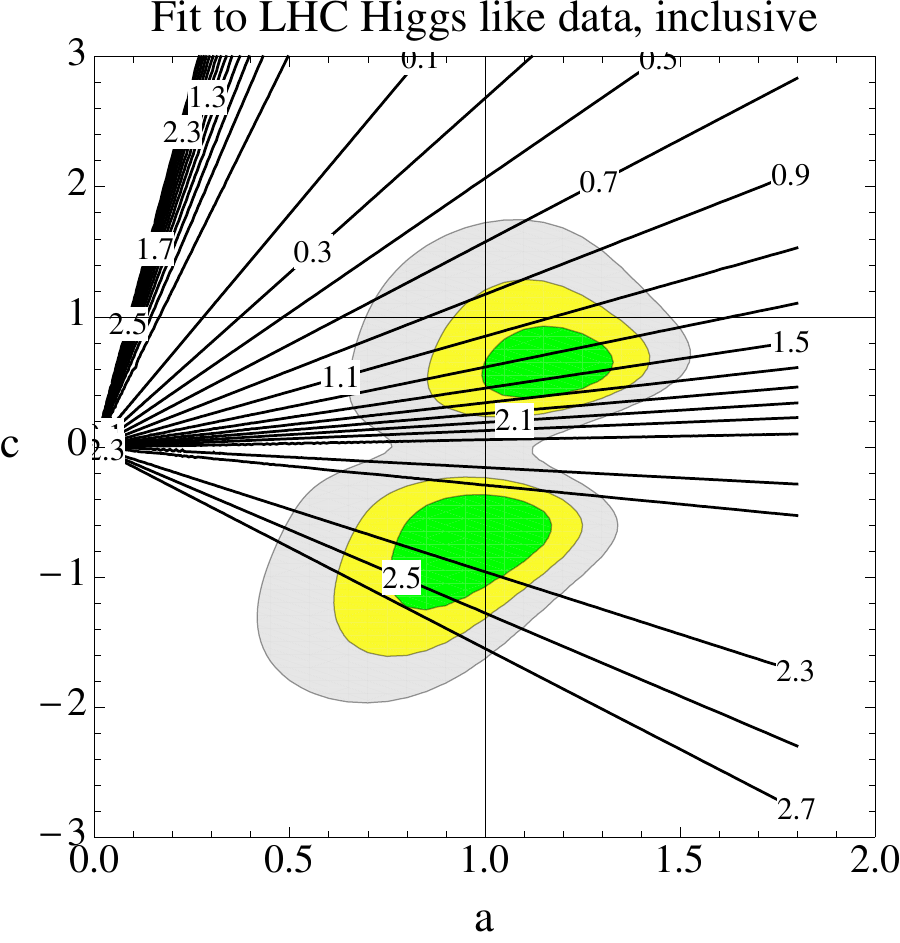}
\includegraphics[width=0.49\textwidth]{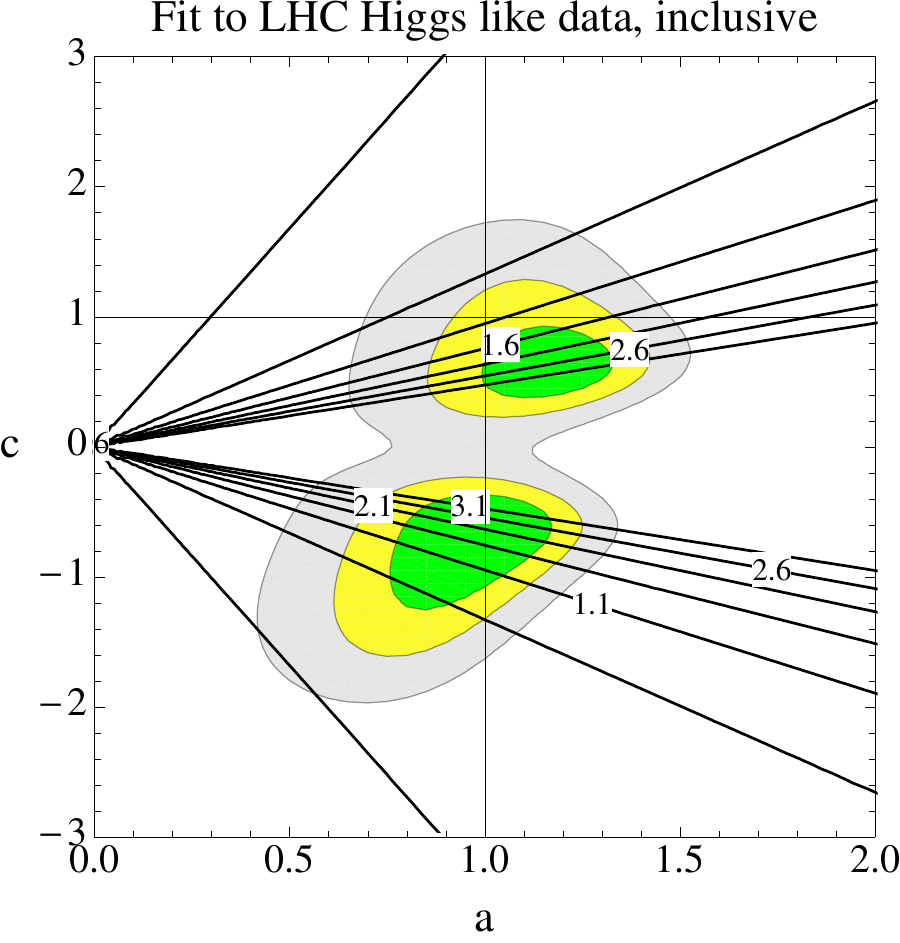}
\caption{Contours of constant ratios of signal production in the
  $(a,c)$ plane overlapped on the best fit region. On the left, the
  ratio $\mu^{\gamma \gamma}/\mu^{ZZ}$ is superimposed for 8  $\rm TeV$ c.m. energy on the best fit regions,
on the right, the ratio $\mu^{\gamma\gamma}_{VBF}/\mu^{\gamma \gamma}$ is superimposed. These observable ratios can resolve the degeneracy of the best fit regions.}\label{Fig.awesomeness}
\end{figure}

Experimental results can be presented in such a manner as to aid in this pursuit. When best fit signal strengths are presented for a common Higgs mass scale, 
correlation coefficients should be supplied as well, which will lead to more accurate fits. The effective
theory approach makes clear that it is instructive to experimentally provide ratios of various best fit signal strengths, so that the
parameters $(a,c)$ can be more precisely determined. In particular, it
would be helpful if the experimental collaborations
provide results that allow the degeneracy in the two fit minima
to be resolved. Ratios of effective signal strengths can
help in resolving this degeneracy.
In Fig.~\ref{Fig.awesomeness} we show the effective production contour ratios for two combinations of extracted best fit signal strengths in the $(a,c)$ plane,
\bea
\frac{\mu^{\gamma \gamma}}{\mu^{ZZ}} = \frac{(\sigma\times BR)^{h\gamma\gamma}[a,c]}{(\sigma\times  BR)^{hZZ}[a,c]} \, \frac{(\sigma\times BR)_{SM}^{hZZ} }{(\sigma\times BR)_{SM}^{h\gamma\gamma}}, 
\eea
\bea
\frac{\mu^{\gamma\gamma}_{VBF}}{\mu^{\gamma \gamma}} = \frac{(\sigma\times BR)^{h\gamma\gamma}_{VBF}[a,c]}{(\sigma\times BR)^{h\gamma\gamma}[a,c]} \, \frac{(\sigma\times BR)_{SM}^{h\gamma\gamma}}{(\sigma\times BR)_{VBF, SM}^{h \gamma \gamma}}. 
\eea
Here we have included superscripts for the various production channels to make explicit the ratios to be constructed.
For example $\sigma^{hZZ}$ means the combination of the production channels discussed in Section~\ref{resultfit} that are included in $ZZ$ signal events.
These ratios are obviously $1$ in the SM.
It is important to note that comparing theoretical and experimental
determinations of such ratios, which include sets of best fit
signal strengths simultaneously, will allow the degeneracy of the
best fit regions to be significantly reduced. Such combinations can
also be experimentally appealing when they allow systematic
uncertainties to be cancelled, such as photon systematic uncertainties in $\mu^{\gamma\gamma}_{VBF}/\mu^{\gamma \gamma}$. 

\section{Conclusions}

We have examined the current LHC data in an effective theory to
determine to what degree the SM Higgs hypothesis is emerging from the data.
To this end we have performed global fits of best fit signal strengths
and exclusion regions, taking into account current data. The SM Higgs hypothesis
turns out to be consistent with the data at the $82 \, \%$
CL. In our global fits we find that there are two best fit regions. We have determined experimentally accessible ratios
of best fit signal strengths for a specific Higgs mass value that will allow the degeneracy in the best fit regions to be significantly reduced with
sufficient data collected at  8 $\rm TeV$ c.m. energy.

\appendix

\section{Fitting Assumptions} \label{assumefit}

In order to perform the fit we have made several assumptions, some of which can clearly be relaxed with more input from experimental collaborations.
In this Appendix we discuss these assumptions in more detail.

\begin{itemize}
\item[(i)] We have assumed that the excess events reported in Table I, which have various best fit mass values, correspond to the same underlying physics,
which we are assuming can be characterized by the Lagrangian in
Eq.~(\ref{eq:efflag}). We have argued that symmetry considerations
lead us to consider this Lagrangian but the degree to which the reported best fit values
of $\mu^i$ can be associated with a particular mass scale is less clear. The mass resolution in all the various channels is larger than the spread of the Higgs masses listed in Table I, which differ at most by $2.5 \, {\rm GeV}$. However,
reported best fit values of $\mu^i$ depend on the assumed $m_h$ value.
Considering the marginal statistical significance of the various channels, background fluctuations convoluted with signal events can be expected to shift the best fit value of the mass. Characterizing precisely the
effect of such fluctuations is challenging. It is reasonable, however, to 
consider such effects to support our association of the excess with a
common mass scale, which we choose to be $124 \, {\rm GeV}. $
\item[(ii)] We have assumed that the given best fit values of
  $\mu^i$ in the various channels are uncorrelated to one another, neglecting both theory and experimental correlations. This is due to the lack of correlations reported by the experimental
collaborations. This assumption is most easy to address with further
experimental input. Our fitting procedure can be easily modified
to include such correlation coefficients as off-diagonal elements in the
covariance matrix. The signals naively expected to be most strongly
correlated  are the $\gamma \gamma$ events. We have studied the effects of
correlations on the fit by introducing pseudo-correlations
through a correlation coefficient of size $0.5$ between all
 $\gamma \gamma$ data. The fit results are robust against
such pseudo-correlations and we still find two best fit regions
with a similar parameter space as without correlations
taken into account. We have also examined the robustness of the fit
results against a set of other pseudo-correlations randomly chosen and
assigned as off-diagonal elements in the 
covariance matrix. Again we find robust fit results when globally fitting the data.  
\item[(iii)] We have assumed that the effect of rescaling the cross
  section and the branching ratios can be directly associated with a
  rescaling of $\mu^i$, neglecting the effect that
    rescaling the various channels modifies the differential
  distributions. This assumption can fail in the presence of higher
  dimensional operators. The modification of the differential
  distributions will affect $\mu^i$ due to
    experimental selection cuts modifying the shape of the
  differential distributions.
However, we consider this effect to be subdominant to the effect
we have incorporated by fitting to an unfixed $(a,c)$ with production cross sections
and branching ratios modified accordingly. Furthermore, we rescaled the various branching ratios
according to the couplings included in the leading order formulae of the
decay widths. This procedure is consistent when including only QCD
corrections to the decay widths, however when including higher order EW
corrections further effects due to $a, c$ differing from 1 are neglected. Again, this effect is subdominant to the effects that we
have retained.
\item[(iv)] We have also neglected, in using $\mu^i$, that this result is determined assuming the SM in the combination of sub-channels, leading to the reported  $\mu^i$
ratio. We also expect this effect to be subdominant to the effects
that we have retained. This is in particular the case if the sub-channel combination is dominated by a particular final state. The analyses that use sub-channel combinations
are the $\rm W \, W$, $\rm Z Z$, $\tau \tau$ channels. We have tested
the robustness of the fit by fitting with two procedures shown in 
Fig.~\ref{Fig.global}. In the figure labeled `$gg$ only', we have only
included the dominant production cross section leading to the 
various final states (usually $\sigma_{gg \rightarrow h}$
production). We also have performed a fit where all production
processed are included. The latter results are labeled as `inclusive' in Fig.~\ref{Fig.global}. We again find that the fit is robust against such differences.
\item[(v)] We have neglected the effects of higher dimensional
  operators in the fit. This assumption can be justified in composite Higgs models, where the
  shift symmetry of the pseudo-Goldstone Higgs suppresses such operators,
and the coefficients we have retained are more sensitive to the underlying strong dynamics. 
\item[(vi)] The SM inputs used in the fit are $m_h = 124 \, {\rm GeV}$, $m_W = 80.398 \, {\rm GeV}$, $m_Z = 91.1876 \, {\rm GeV}$, $m_t = 172.5 \, {\rm GeV}$ and $\sin^2(\theta_W)= 0.223$.
\end{itemize}

It is clearly of interest to further examine the effect of these assumptions, relaxing them whenever possible theoretically and to perform more refined analyses of this form when further experimental input is available. 

\section{Moriond 2012 Update} \label{assumefit}
In this section we present updated results including the data presented at Moriond 2012 \cite{moriond}.
The most significant change in the data that was used in Table I is an update to the ATLAS measurement of 
$pp \rightarrow  W \, W^\star \rightarrow \ell^+ \, \nu  \, \ell^- \, \bar{\nu}$. In addition, ATLAS reported best fit
signal strengths in the $pp \rightarrow  b \, \bar{b}$ and $pp \rightarrow  \tau \, \bar{\tau}$ channels while CDF and ${\rm D}0 \! \! \! /$ reported a broad excess in
$p\bar{p} \rightarrow  b \, \bar{b}$ events. Further, CMS has now also supplied best fit signal strengths as a function of $m_h$, allowing various Higgs mass hypotheses to be fit to. 
In this Appendix we include these experimental results in our fit and supply supplementary plots for various Higgs masses (refining also our determination of the 95\% C.L. exclusion limits).
\begin{figure}[tb]
\includegraphics[width=0.45\textwidth]{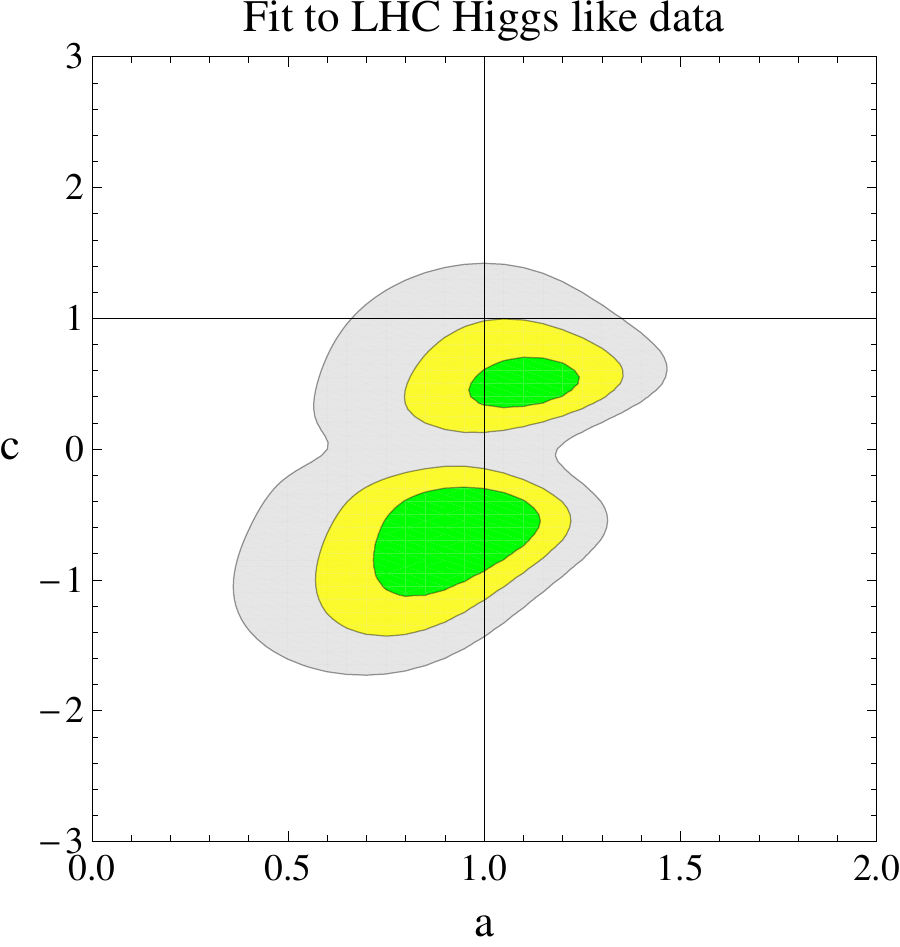}
\includegraphics[width=0.45\textwidth]{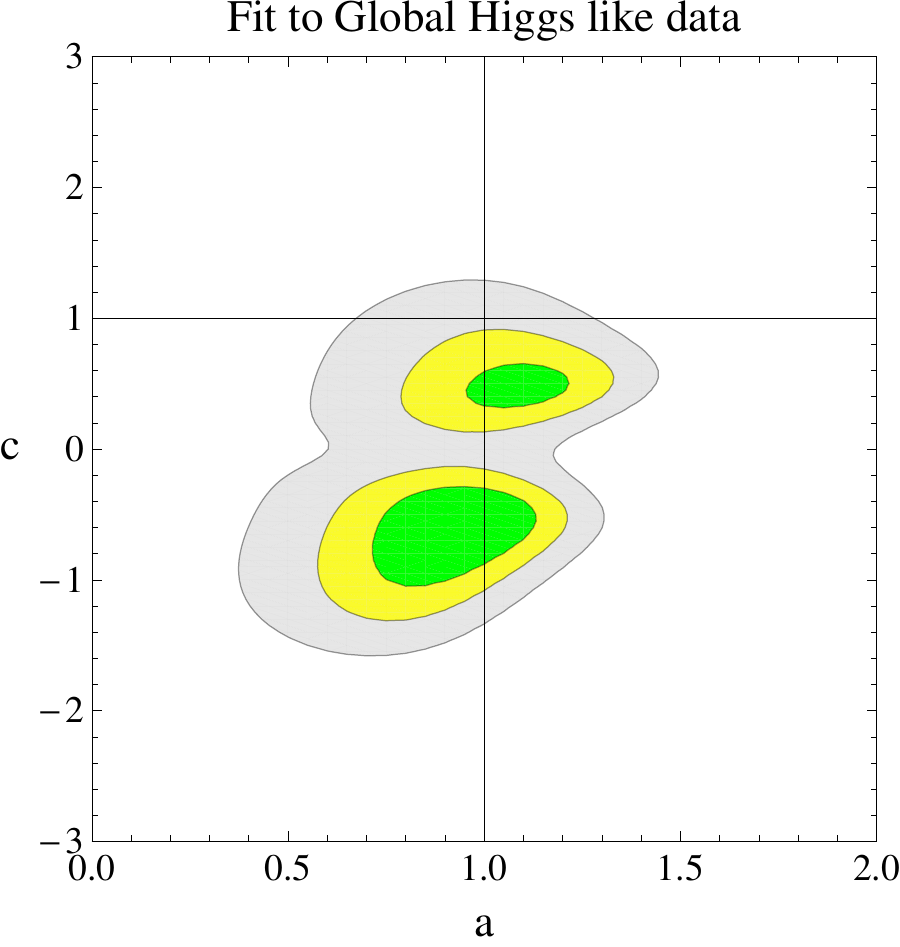}
\caption{Update to the global fit. To the left we have only updated the $WW$ ATLAS signal in Table I to the value in Table II. On the right we have also added the ATLAS $\tau \tau$ and
Tevatron data on $p\bar{p} \rightarrow  b \, \bar{b}$ and  $p\bar{p} \rightarrow W^+ \, W^-$ as shown in Table II.}
\end{figure}
\begin{table} 
\setlength{\tabcolsep}{5pt}
\center
\begin{tabular}{c|c|c} 
\hline \hline 
Channel [Exp] &  $m_h [{\rm GeV}]$ & $\mu$ ($\mu_L$) 
\\
\hline
$pp \rightarrow  W \, W^\star \rightarrow \ell^+ \, \nu  \, \ell^- \, \bar{\nu} \,\, [{\rm ATLAS}] $ & $126$ & $0.2^{+0.6}_{-0.7}$ \, (1.3) \\
$pp \rightarrow  b \, \bar{b} \,\, [{\rm ATLAS}] $ &   $124 $ & $-0.8^{+1.7}_{-1.7} $ \, (3.5)  \\
$pp \rightarrow  \tau \, \bar{\tau} \,\, [{\rm ATLAS}]$ &  $124 $  & $-0.1^{+1.7}_{-1.7}$ \,  (3.4) \\
$p\bar{p} \rightarrow  b \, \bar{b} \,\, [{\rm CDF \& D 0\! \! \! /}]$ &  $125 $ & $2.0^{+0.8}_{-0.7}$ \, (3.2) \\
$p\bar{p} \rightarrow W^+ \, W^- \,\, [{\rm CDF \& D 0\! \! \! /}]$ &  $125 $ & $0.03^{+1.22}_{-0.03}$ \, (2.4) \\
\hline \hline
\end{tabular}
\caption{Summary table of reported signatures with events related to the Higgs mass scale of interest ($m_h\simeq 124$~GeV) where excess events have been reported. Moriond 2012 update with new numbers
to supplement (or replace) entries in Table I.}
\label{table:tcorrections} \vspace{-0.35cm}
\end{table}

The updated data that we supplement Table I with is given in Table II. Due to an apparent inconsistency in the ATLAS best fit signal strength plot for $pp \rightarrow  b \, \bar{b}$
and the corresponding ATLAS $\rm CL_s$ limit plot we do not use the $b \, \bar{b}$ best fit signal strength value in the combined fit.
We show in Fig.~7 the effect of the Moriond 2012 data on our previously reported fit results.

\begin{table} 
\setlength{\tabcolsep}{5pt}
\center
\begin{tabular}{c|c|c|c} 
\hline \hline 
Channel [Exp] &  $\hat{\mu}_{119.5}$ ($\mu_{119.5}^L$) & $\hat{\mu}_{124}$ ($\mu_{124}^L$)  & $ \hat{\mu}_{125}$ \, ($\mu_{125}^L$)
\\
\hline
$pp \rightarrow \gamma \, \gamma \,\, [{\rm ATLAS}]$  & $0.0^{+0.6}_{-0.8} \, \,(1.5) $ & $0.8^{+0.8}_{-0.7} \, \, (2.6) $ & $1.6^{+0.9}_{-0.8}\,\, (3.9) $ 
\\
$pp \rightarrow  Z \, Z^\star \rightarrow \ell^+ \, \ell^- \, \ell^+ \, \ell^- \,\, [{\rm ATLAS}] $ & $-0.5^{+1}$ \, (5.1) & $1.6^{+1.4}_{-0.8}$ \, (4.7) &  $1.4^{+1.3}_{-0.8}$ \, (4.1) 
\\
$pp \rightarrow  W \, W^\star \rightarrow \ell^+ \, \nu  \, \ell^- \, \bar{\nu} \,\, [{\rm ATLAS}] $ &  $0.0^{+1.2}_{-1.3}$ \, (2.4) & $0.1^{+0.7}_{-0.7}$ \, (1.6) &  $0.1^{+0.7}_{-0.6}$ \, (1.4) 
\\
$pp \rightarrow  \gamma \, \gamma \,\, [{\rm CMS}]$ & $-1.1^{+0.6}_{-0.6}$ \, (1.3) & $1.5^{+0.7}_{-0.7}$ \, (3.5) & $1.6^{+0.7}_{-0.6}$ \,  (3.0)  
\\
$pp \rightarrow  Z \, Z^\star \rightarrow \ell^+ \, \ell^- \, \ell^+ \, \ell^- \,\, [{\rm CMS}] $ &  $2.0^{+1.6}_{-1.1}$ \, (5.2) & $0.5^{+1.1}_{-0.7}$ \, (2.7) & $0.6^{+0.9}_{-0.6}$ \, (2.5)
\\
$pp \rightarrow  W \, W^\star \rightarrow \ell^+ \, \nu  \, \ell^- \, \bar{\nu} \,\, [{\rm CMS}] $ & $0.9^{+0.8}_{-0.7}$ \, (2.5)& $0.6^{+0.7}_{-0.7}$ \, (1.8) & $0.4^{+0.6}_{-0.6}$ \, (1.5) \\
$pp \rightarrow  b \, \bar{b} \,\, [{\rm CMS}] $ &  $0.4^{+1.8}_{-1.6}$ \, (4.1) & $1.2^{+1.9}_{-1.8}$ \, (5.0) & $1.2^{+2.1}_{-1.7}$ \, (5.2) 
\\
$pp \rightarrow  \tau \, \bar{\tau} \,\, [{\rm CMS}]$ &  $0.2^{+0.9}_{-1.1}$ \, (3.6)  & $0.4^{+1.0}_{-1.2}$ \, (3.9) & $0.6^{+1.1}_{-1.2}$ \, (4.1)
\\
$pp \rightarrow  \tau \, \bar{\tau} \,\, [{\rm ATLAS}]$ &  $-0.9^{+1.7}_{-1.7}$ \, (2.9)  & $-0.1^{+1.7}_{-1.7}$ \, (3.4) & $0.1^{+1.7}_{-1.8}$ \, (3.5)
\\
$p\bar{p} \rightarrow  b \, \bar{b} \,\, [{\rm CDF \& D 0\! \! \! /}]$ & $1.5^{+0.6}_{-0.5}$ \, (2.5) & $1.9^{+0.8}_{-0.6}$ \, (3.1) & $2.0^{+0.8}_{-0.7}$ \, (3.2)\\
$p\bar{p} \rightarrow W^+ \, W^- \,\, [{\rm CDF \& D 0\! \! \! /}]$ & $1.63 ^{+1.46 }_{-1.12}$ \, (4.5) & $0.03 ^{+1.22 }_{-0.03}$ \, (2.4) & $0.03 ^{+1.22 }_{-0.03}$ \, (2.4) \\
\hline \hline
\end{tabular}
\caption{Summary table of reported best fit signal strengths for various Higgs mass values. We note that the asymmetric nature of the error band for the best fit signal strength values reported for the $pp \rightarrow  Z \, Z^\star \rightarrow \ell^+ \, \ell^- \, \ell^+ \, \ell^-$ signal by ALTAS is curious. We use this data in the fit. For $m_h = 125 \, {\rm GeV}$ we use the public results presented at Moriond 2012 that splits the $\gamma \, \gamma$ signal events into
four classes (that are not identical to the classes used in the body of the paper) as well as the (dominantly) VBF induced photon events instead of the global photon value in the Table. The non VBF di-photon events we rescale inclusively. 
The data used for the category zero to four photons is given (in order) for $\mu_{125}$
by $2.1^{+2.0}_{-1.6}$, $0.6^{+1.0}_{-0.3}$, $2.2^{+1.4}_{-1.4}$, $0.5^{+1.8}_{-1.7}$, while the VBF induced photon result is $3.6^{+2.2}_{-1.6}$.}
\label{table:tcorrections} \vspace{-0.35cm}
\end{table}

\begin{figure}[tb]
\includegraphics[width=0.45\textwidth]{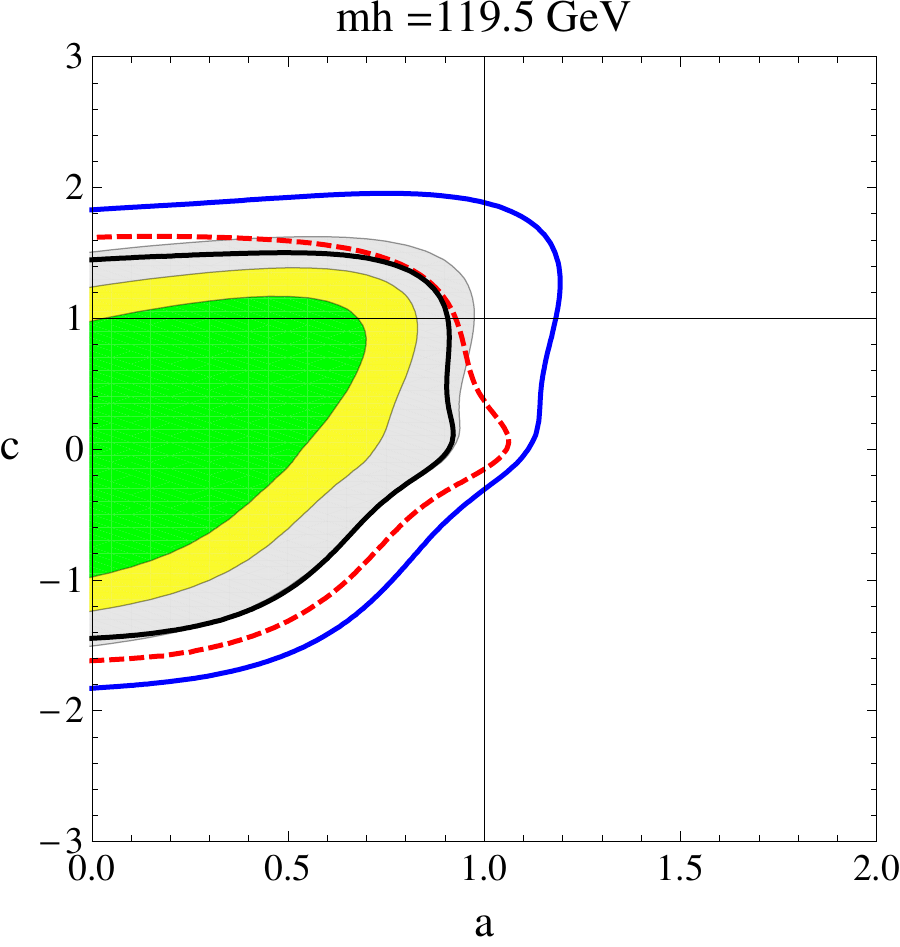}
\includegraphics[width=0.45\textwidth]{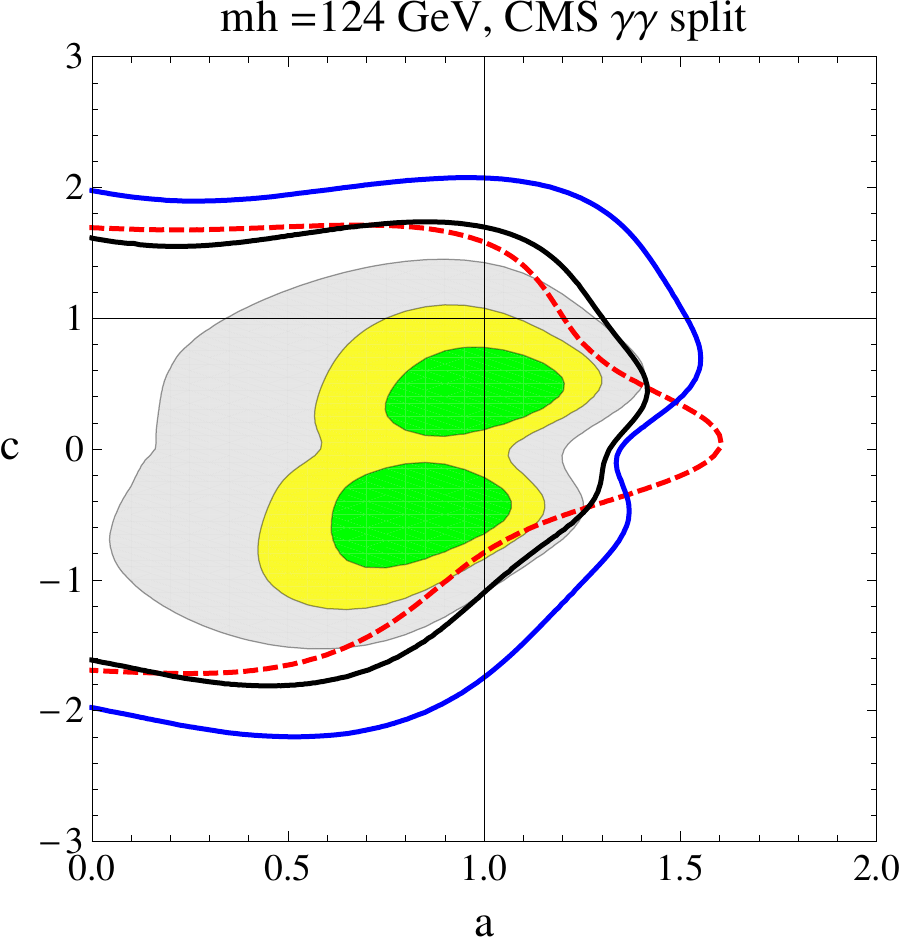}
\includegraphics[width=0.45\textwidth]{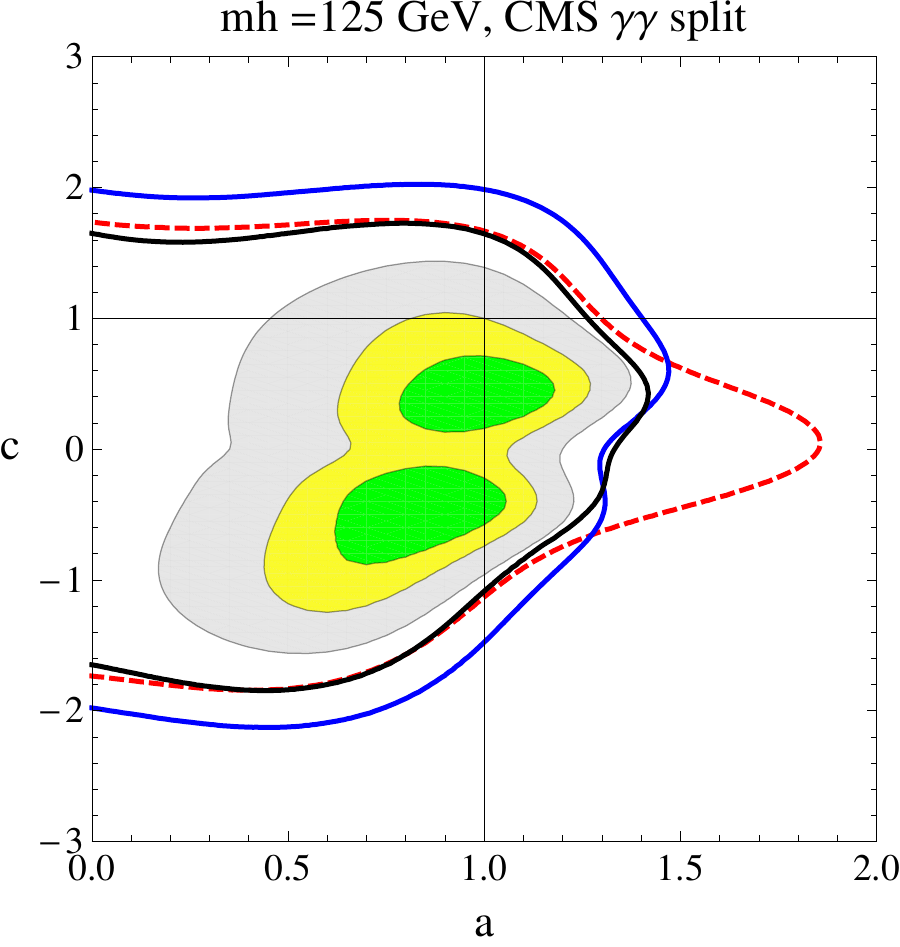}
\includegraphics[width=0.45\textwidth]{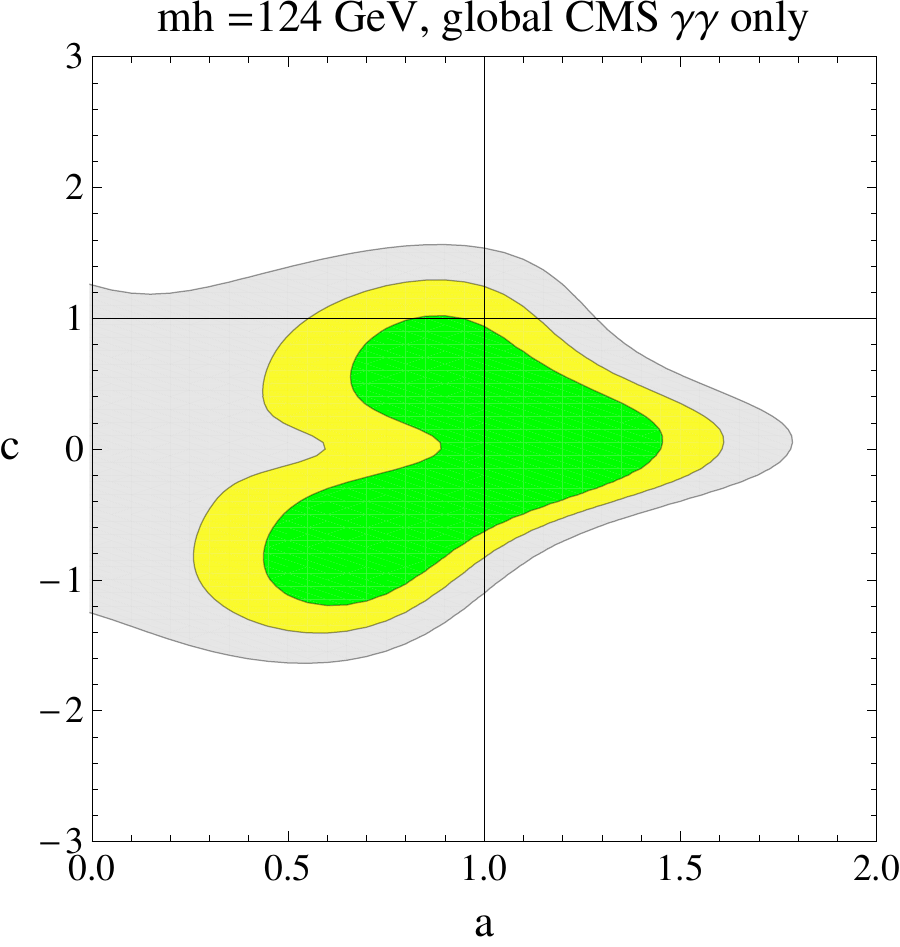}
\caption{Fitting various mass hypothesis using the data from Table III and its caption. The red dashed line is the ATLAS exclusion limit as described in the text, the blue solid line is the CMS limit and the combined
CMS and ATLAS limit is included as a black solid line.}
\end{figure}

We also show joint fits for the Higgs mass values $m_h = 119.5, 124,125 \, {\rm GeV}$ where we have taken the experimentally reported $\hat{\mu}$ and the corresponding theory predictions at a common $m_h$ due to 
the release of the required data by CMS, after version one of this paper. The data we use is given in Table III for each particular Higgs mass value chosen. These results are shown for various masses in 
Fig.~8. Examining the results, one clearly sees that the excess at $119.5 \, {\rm GeV}$ can be distinguished globally to be a likely statistical fluctuation compared 
to the global fit to the excess of events around $\sim 124-126 \, {\rm GeV}$. For the three mass values, the best fit points $(m_h,a,c,\chi^2)$ are $(119.5, 0.28,0.49, 9.8)$, $(124, 0.87,-0.43, 4.1)$ and
$(125, 0.87,-0.42, 4.6)$, where for the latter two masses we use the CMS di-photon data resolved into subclasses.
Fitting to $m_h = 124 \, {\rm GeV}$ with global di-photon best fit data only, compared to di-photon data split into subclasses, shows that the discrimination on the parameter space offered
by separately reporting the VBF induced $\gamma \gamma \, jj$ signal is important. We also encourage the experimental collaborations to report the degree of contamination
of this signal with $gg$ initial state Higgs events to enable a consistent treatment of the reported best fit signal stengths. One can demonstrate how the VBF signal interpolates between the two results shown in Fig.~8 by adding in contamination
due to $\sigma(gg \rightarrow h)$ events with our consistent rescaling procedure. One finds the series of plots shown in Fig.~9 for various degrees of contamination. Note that CMS also reports $W^+ \, W^- \, jj$ events
which offer a similar discrimination of the parameter space as the $\gamma \gamma \, jj$ signal in principle. However, again contamination
due to $\sigma(gg \rightarrow h)$ events will exist and is not reported by the collaborations. We do not use
this data as a separate channel at present. 
\begin{figure}[tb]
\includegraphics[width=0.45\textwidth]{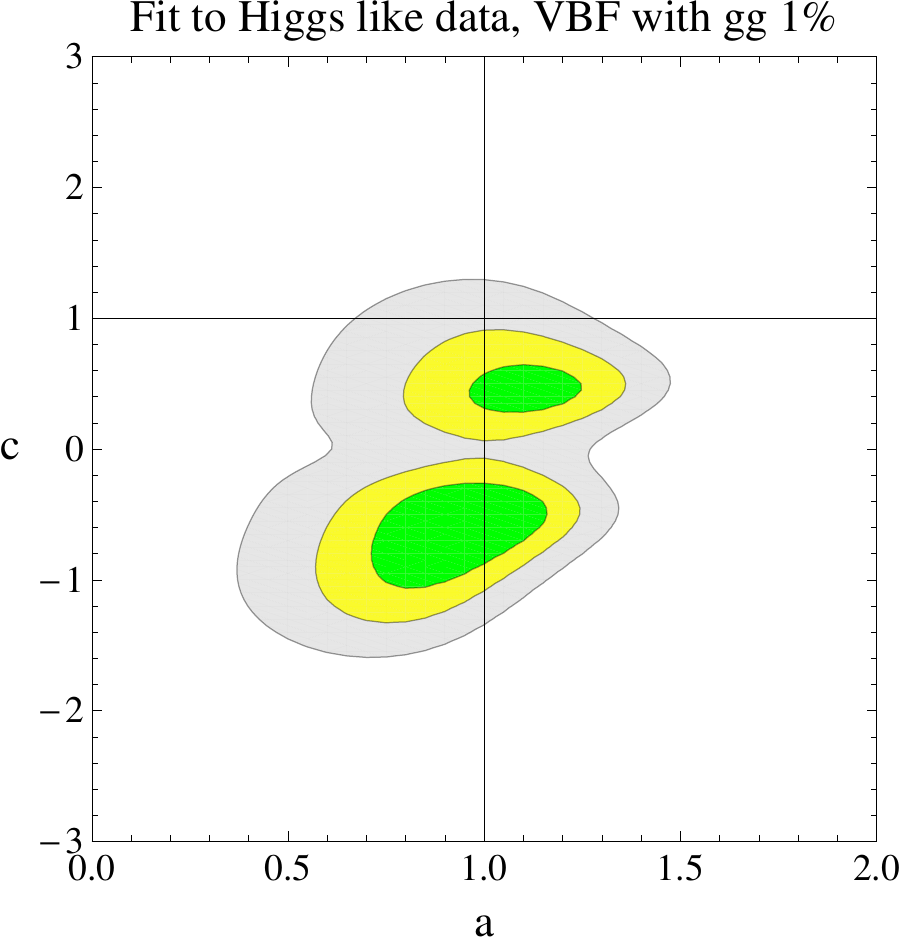}
\includegraphics[width=0.45\textwidth]{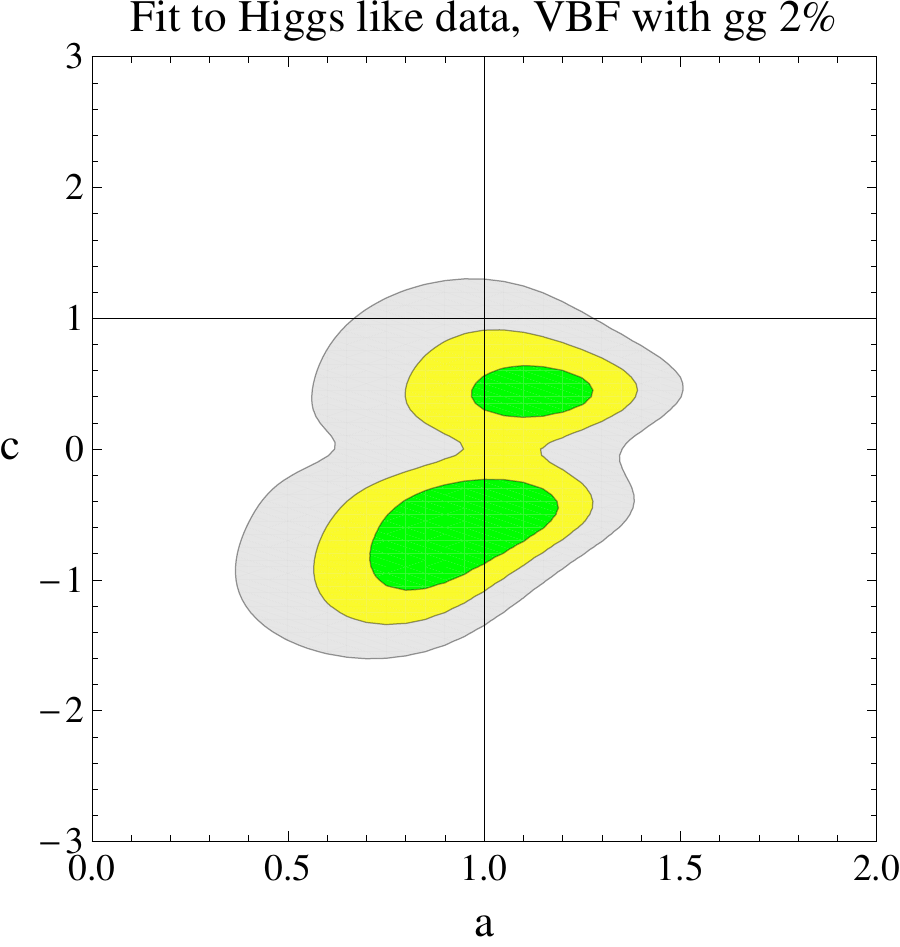}
\includegraphics[width=0.45\textwidth]{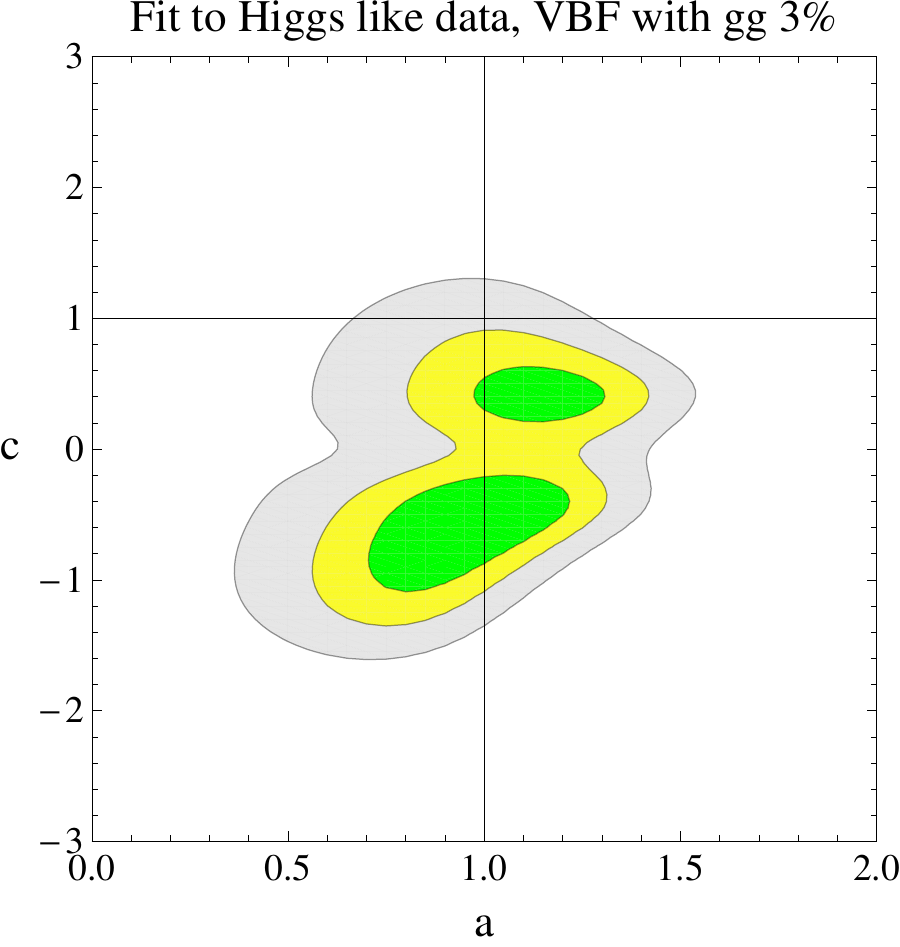}
\includegraphics[width=0.45\textwidth]{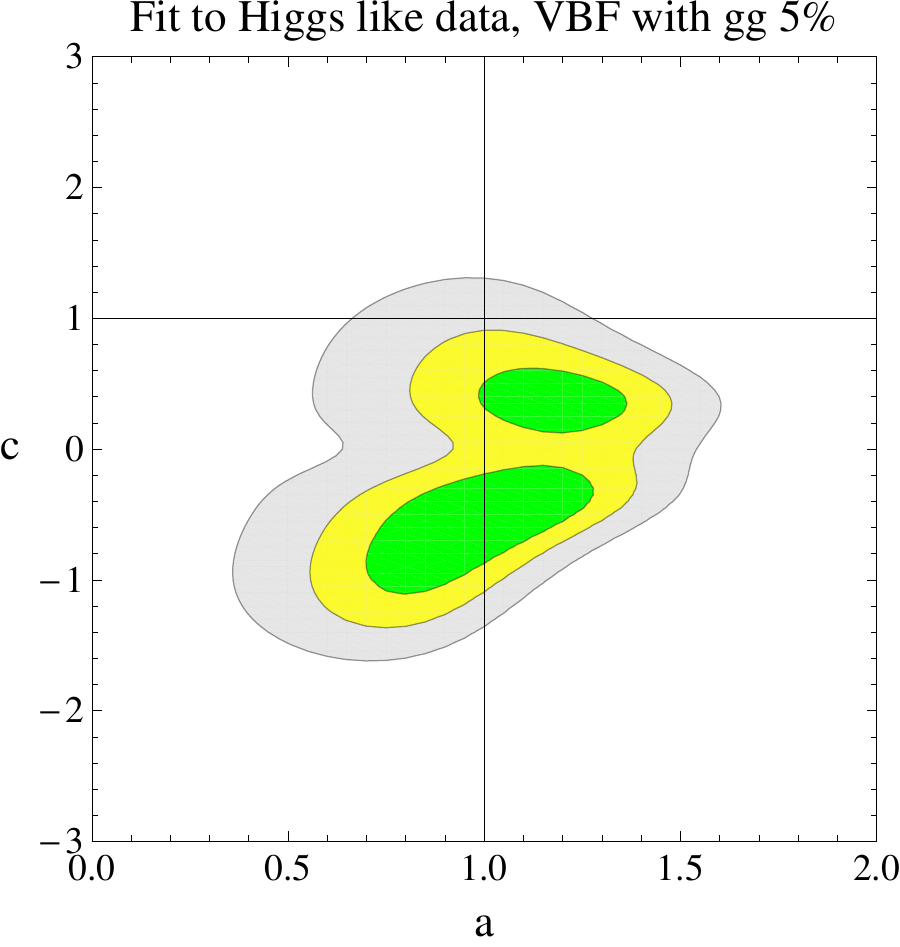}
\caption{Various degrees of contamination of the VBF signal with $gg \rightarrow h$ events.}
\end{figure}
The update to the data has a small effect on the ${\rm CL}$ of the SM Higgs hypothesis compared to the best fit value of the current data. Assuming no contamination due to gg for VBF events one finds that our previously reported
global fit with the Moriond 2012 data update (but without correcting to a single Higgs mass value in the experimental best fit signal strengths) has the SM hypothesis residing on a $94\, \% \,{\rm CL}$ curve around the best fit value of $(a,c)$. 
Assuming a $3 \%$ contamination of the VBF events due to gg,
the SM hypothesis remains consistent with the data at $93\, \% \, {\rm CL}$ compared to the best fit value.  For a direct comparison, the Fermiophobic scenario with $a = 1, c = 0$ is consistent with the data at $96 \, \% {\rm CL}$ for the same global fit.
When a $3\%$ contamination due to gg events for the VBF diphoton signal of CMS is assumed, the same Fermiophobic scenario  is consistent with the data at the $88 \, \% {\rm CL}$.
We do not consider a Fermiophobic scenario to be favoured by the global data or the pattern of deviations from the SM in the current data set. With such marginal signal events, statistical fluctuations in the data are still present
affecting the pattern of deviations.

\begin{figure}[tb]
\includegraphics[width=0.49\textwidth]{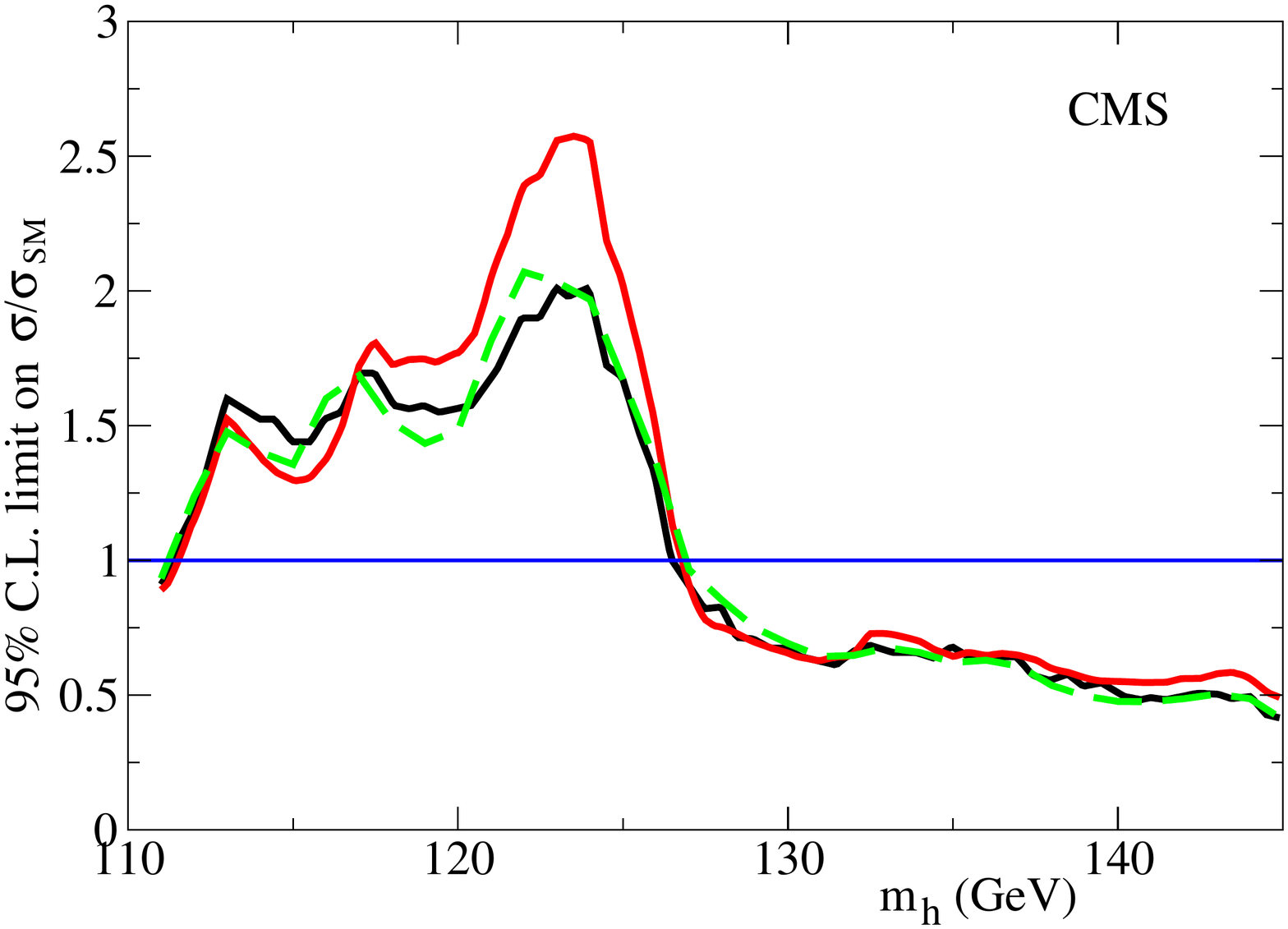}
\includegraphics[width=0.49\textwidth]{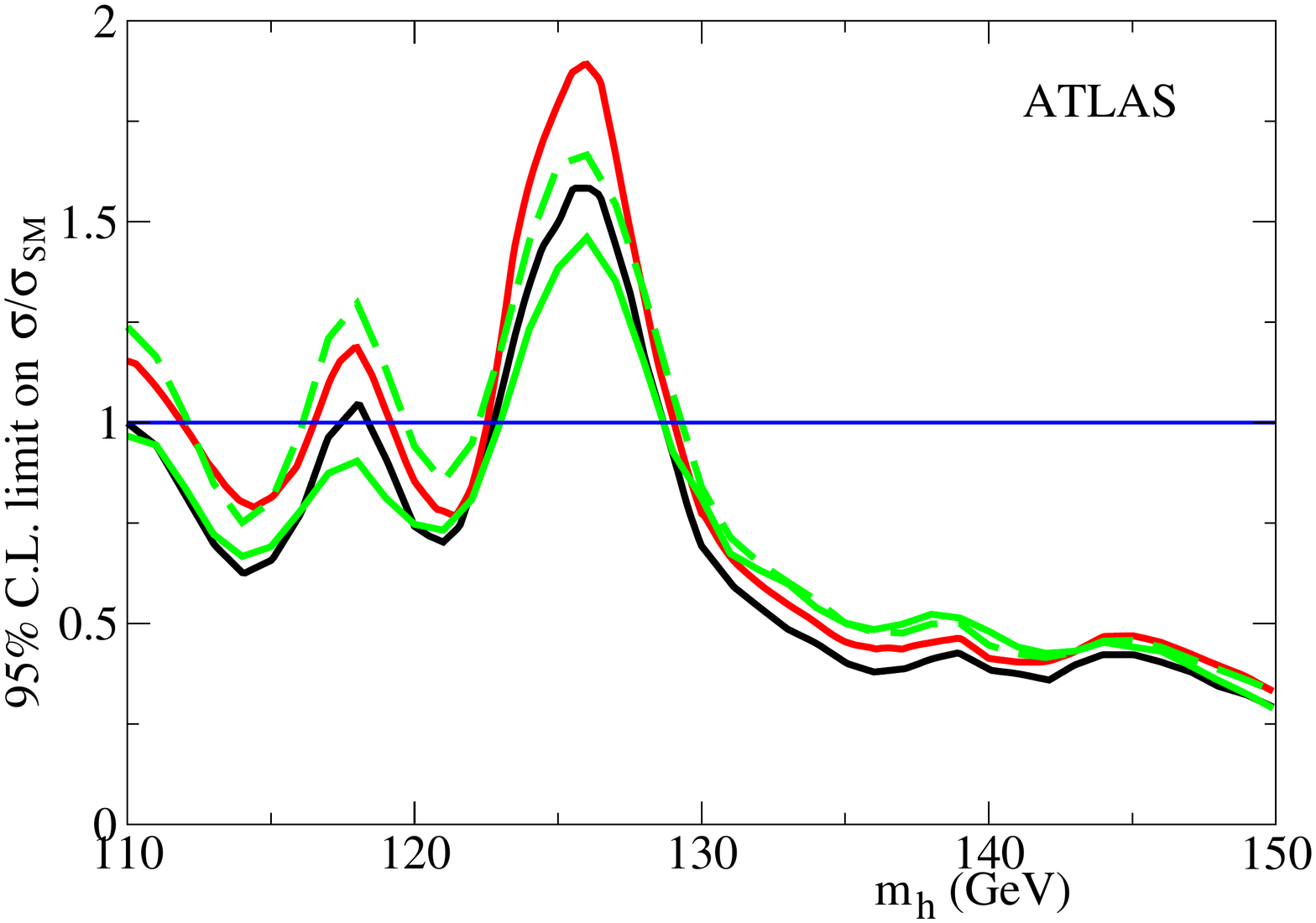}
\caption{95\% C.L. limits on $\sigma/\sigma_{SM}$ for CMS (left) and ATLAS (right). The official curve is the black solid line. The red curve is obtained from the simple approximation of Eqn.(\ref{applim}). The green curves are more precise determinations of the limit as explained in the text.}\label{fig:limitscomp}
\end{figure}

Finally, the 95\% CL exclusion curves from ATLAS and CMS in the plots of this Appendix have been
determined using a more precise method than Eq.~(\ref{applim}). In the same spirit of Ref.~\cite{Azatov:2012bz},
for each individual search channel $i$, we have first approximated the corresponding probability density
function of the signal strength parameter $\mu$ by a Gaussian $p_i(\mu)\propto
\mathrm{Exp}[-(\mu-{\bar\mu}^i)^2/(2\sigma_{\mathrm{obs},i}^2)]$. We have obtained the quantities 
${\bar\mu}^i$ and $\sigma_{\mathrm{obs},i}$ trying to get the best approximation to the reported
95\% CL in that channel, $\mu_{L}^i$ (obtained from the equation $\int_0^{\mu_{L}^i}p_i(\mu)d\mu=0.95$).
In the case of CMS channels, we use the approximation of Ref.~\cite{Azatov:2012bz} (which uses 
$\sigma_{\mathrm{obs},i}\simeq \sigma_{\mathrm{exp},i}=\mu_{L,exp}^i/1.96$ and obtains $\bar\mu$ by solving
the equation that determines $\mu_{L}^i$). In the case of ATLAS data, we find better agreement with
the reported limits by directly using $\bar\mu^i=\hat\mu^i$ and $\sigma_{\mathrm{obs},i}$ as provided\footnote{This is not the case of the $b\bar{b}$ channel, where we find a discrepancy between the limit derived from 
$\hat\mu^i$ and $\sigma_{\mathrm{obs},i}$ and the reported limit. We therefore do not use this channel to extract 
the combined limit in the $(a,c)$ plane. Although this is a subdominant channel, the impact of such discrepancy on the SM exclusion around 
$115-120$ GeV is not negligible. In the case of the $Z\rightarrow 4l$ channel, for some ranges of Higgs masses
the reported (negative) value of $\hat\mu$ is hitting a boundary and we could not use it to reproduce well the experimental limit. In
these cases we use the same approximation as for the CMS channels.}. To illustrate the precision of our approximations to the exclusion limits we compare them with the official 95\% CL limit on $\sigma/\sigma_{SM}$ in Fig.~\ref{fig:limitscomp}. In the left (right) panel we show the CMS (ATLAS) limit, with the official curve in black. The red curve is the simple approximation of Eq.~($\ref{applim}$) and the green curves are more precise determinations of the limit as explained above. The dashed green line corresponds to
an approximate determination of ${\hat\mu}^i$ and $\sigma_{\mathrm{obs},i}$ as in Ref.~\cite{Azatov:2012bz}.
The solid green line (only shown for ATLAS) uses  ${\bar\mu}^i={\hat\mu}^i$ and $\sigma_{\mathrm{obs},i}$ as provided.

\subsection*{Acknowledgments}
As the draft of this work was being finalized two analyses with overlap to some of the work presented here
appeared \cite{Carmi:2012yp,Azatov:2012bz}. We thank Ricardo Gon\c{c}alo, Aurelio Juste,  Michael Spira, Wade Fisher, Joey Huston,
Vivek Sharma and Josh Bendavid for helpful communication on related theory and data.
This work has been partly
supported by the European Commission under the contract ERC advanced
grant 226371 �MassTeV�, the contract PITN-GA-2009-237920 �UNILHC�, and
the contract MRTN-CT-2006-035863 �ForcesUniverse�, as well as by the
Spanish Consolider Ingenio 2010 Programme CPAN (CSD2007-00042) and the
Spanish Ministry MICNN under contract FPA2010-17747 and
FPA2008-01430. MM is supported by the DFG SFB/TR9 Computational Particle Physics.

\end{document}